\documentclass[prd,aps,
twocolumn,
preprintnumbers,
notitlepage,
nofootinbib]{revtex4}

\usepackage{graphics,graphicx,epsfig,epsf,amsmath,amssymb,amsfonts}
\usepackage{float}
\usepackage{multirow}
\usepackage{array}
\usepackage{dcolumn}
\usepackage{bm}
\usepackage{hyperref}
\usepackage[usenames]{color}
\begin{document}
\newcommand{\vecc}[1]{\mbox{\boldmath $#1$}}
\newcommand{\no}{№ } 
\newcommand\Tr{\textmd{Tr\,}}
\def\be{\begin{equation}}
\def\ee{\end{equation}}
\def\beq{\begin{equation}}
\def\eeq{\end{equation}}
\def\GEV{\mbox{GeV}}
\def\ba{\begin{eqnarray}}
\def\ea{\end{eqnarray}}
\def\bc{\begin{center}} 
\def\ec{\end{center}}   
\def\nn{\nonumber}
\def\ni{\noindent}
\definecolor{navyblue}{rgb}{0.0, 0.0, 1.0}
\definecolor{darkred}{rgb}{1.0, 0.0, 0.0}
\definecolor{modRed}{RGB}{230,82,90}
\definecolor{red(ryb)}{rgb}{1.0, 0.15, 0.07}

\title{
    Polarization effects in the elastic  $e \vec p \to \vec e  p$ and $\vec e \vec p \to  e  p$\\
    processes in the case of parallel spins}

\author{M.V. Galynskii}
\email{galynski@sosny.bas-net.by}
\affiliation{
Joint Institute for Power and Nuclear Research -- Sosny,
National Academy of Sciences of Belarus, 220109 Minsk, Belarus}

\author{V.V. Bytev}
\email{bvv@jinr.ru}
\affiliation{
Joint Institute for Nuclear Research, 141980 Dubna, Russia}

\author{V.M. Galynsky}
\email{galynsky@bsu.by}
\affiliation{Belarusian State University, Minsk 220030, Belarus}

\begin{abstract}
In the one-photon exchange approximation, we analyze polarization effects
in the elastic $\vec e \vec p \to e p$ and $ e \vec p \to \vec e p$ processes in the case
when the spin quantization axes of a target proton at rest and an incident or scattered
electron are parallel.
To do this, in the kinematics of the SANE Collaboration experiment
[A.~Liyanage {\it et al.}, \href{https://doi.org/10.1103/PhysRevC.101.035206}{Phys.\ Rev.\ C {\bf 101}, 035206 (2020)}]
using the J.~Kelly [\href{https://dx.doi.org/10.1103/PhysRevC.70.068202}{Phys. Rev. C {\bf 70}, 068202 (2004)}]
and [I.~Qattan {\it et al.} \href{https://dx.doi.org/10.1103/PhysRevC.91.065203}{Phys. Rev. C {\bf 91}, 065203 (2015)}]
parametrizations for the Sachs form factor ratio $R\equiv \mu_p G_E/G_M$, a numerical analysis was carried out
of the dependence of the longitudinal polarization degree transferred to the scattered electron
in the $e \vec p \to \vec e p$ process and double spin asymmetry in the $\vec e \vec p \to e p$ process
on the square of the momentum transferred to the proton as well as on the scattering angle of the electron.
It is established that the difference in the longitudinal polarization degree of the scattered
electron in the $ e \vec p \to \vec e p$ process in the case of conservation and violation
of the scaling of the Sachs form factors can reach 70 \%.
This fact can be used to set up polarization experiments of a new type to measure the ratio $R$.
For  double spin asymmetry in the $\vec e \vec p \to e p$ process, the corresponding difference
does not exceed 2.32 \%. This fact means that it's not sensitive to the effects of the Sachs
form factor scaling violation and could be used as test for the $R\approx 1$ equality.
\end{abstract}

\maketitle


\section{Introduction}

Experiments on the study of electric $G_{E}$ and magnetic $G_{M}$ proton form factors, the so-called
Sachs form factors (SFFs), have been performed since the mid-1950s in the elastic process of
electron-proton scattering \cite{Hofstadter1958}. In the case of unpolarized electrons
and protons, all experimental data on the behavior of the SFFs were obtained  with the help of the Rosenbluth
technique (RT) based on the  Rosenbluth formula for the differential cross section
for the $ep \to ep$ process in the rest frame of the initial proton \cite{Rosen}; that is,
\ba
\label{Ros}
\frac{d\sigma} {d\Omega_e}=
\frac{\alpha^2E_2\cos^2(\theta_e/2)}{4E_1^{3}\sin^4(\theta_e/2)}
\frac{1}{1+\tau_p} \left(G_E^{2}
+\frac{\tau_p}{\varepsilon}G_M^{2}\right).
\ea
Here  $\tau_p=Q^2/4m^2, Q^2=4E_1 E_2\sin^2(\theta_e/2)$ is the square of the 4-momentum
transferred to the proton; $m$ is the mass of the proton; $E_1$, $E_2$ are the energies of the initial
and final electrons, $\theta_e$ is the electron scattering angle;
$\varepsilon=[1+2(1+\tau_p)\tan^2(\theta_e/2)]^{-1}$ is the
degree of linear (transverse) polarization of the virtual photon \cite{Dombey,Rekalo74,AR,GL97};
$\alpha=1/137$ is the fine structure constant. Expression (\ref{Ros}) was obtained in the one-photon exchange
(OPE) approximation and the electron mass was set to zero.

With the help of RT, the dipole dependence of the SFFs on the momentum transferred to the proton
square  $Q^2$ in the region $Q^2\leq 6$ $\GEV^2$ was established (see Ref. \cite{ETG15} for
an exhaustive review). As it turned out, these measurements indicate approximate form
factor scaling, i.e. $\mu_p G_E/G_M \approx 1$, where $\mu_p=2.79$
is the magnetic moment of the proton.

Akhiezer and Rekalo \cite{Rekalo74} proposed a method for measuring the $R \equiv\mu_p G_E/G_M$ ratio
based on the phenomenon of polarization transfer from the initial electron to the final
proton in the $\vec e p\to e\vec p$ process (later this method was generalized in Ref.
\cite{Miller2015}). Precision JLab experiments \cite{Jones00,Gay01,Gay02},
using this method, found a fairly rapid decrease in the ratio of $R$ with an increase in $Q^2$,
which indicates the violation of the dipole dependence over transferred
momentum square $Q^2$. In the range $0.4~\GEV^2 \leqslant Q^2 \leqslant 5.6~\GEV^2$, as
it turned out, this decrease is linear. Next, more accurate measurements of the ratio $R$
carried out in \cite{Pun05,Puckett10,Puckett12,Puckett17,Qattan2005}
in a wide area in $Q^2$ up to $8.5~\GEV^2$ using both the Akhiezer -- Rekalo (AR) method \cite{Rekalo74} and
the RT \cite{Qattan2005}, only confirmed the discrepancy of the results.

In the SANE Collaboration experiment \cite{Liyanage2020}, the values of $R$ 
has been measured for the elastic $\vec e \vec p \to e p$ process
by the third method \cite{Donnelly1986} using 
double spin asymmetry for target spin orientation aligned nearly perpendicular to the
beam momentum direction 
in the case, when the electron beam and the proton target are partially polarized.
The degree of polarization of the proton target was $P_t=(70 \pm 5)$\%. The experiment was performed
at two electron beam energies $E_1$, 4.725 GeV and 5.895 GeV, and two $Q^2$ values,
2.06 GeV$^2$ and 5.66 GeV$^2$. The extracted values of $R$ in \cite{Liyanage2020} are consistent with the
results in Refs. \cite{Jones00,Gay01,Gay02,Pun05,Puckett10,Puckett12,Puckett17}.

Currently, the most precise 
measurements of the proton FFs at low momentum transfer, and its charge and magnetic
radii were performed by the A1 Collaboration at MAMI, Mainz \cite{Bernauer1,Bernauer2,Bernauer3}.
Cross-sections were measured at 1422 kinematic settings covering a $Q^2$ range
from 0.004 to 1.0 GeV$^2$ with average point-to-point systematic error of 0.37\% \cite{Bernauer1,Bernauer2,Bernauer3}.
With this large data set, the authors extracted $G_E$ and $G_M$ by a direct fit of
form factor models to cross-section data rather than the traditional Rosenbluth
separation technique. The results of the fits  reconcile with  a classic
Rosenbluth separation within error estimations.


Thus, while the Rosenbluth data are compatible with the scaling relation prediction, polarized
experiments yield data with a linear, downward trend. The most commonly proposed explanation for this
discrepancy are ``hard'' two-photon exchange (TPE) contributions beyond the standard radiative
corrections to OPE \cite{Arrington:2011dn, Punjabi2015, Afanasev2017}.
Note  that the recent TPE-experiments \cite{Rachek:2014fam, OLYMPUS:2016gso, CLAS:2013mza, CLAS:2014xso, CLAS:2016fvy}
show little evidence for significant contributions beyond OPE
up to $Q^2 \approx 2.3$ GeV$^2$ \cite{Bernauer2024}.
To determine whether ``hard'' TPE contributions could explain the form factor discrepancy
one needs new measurements at higher $Q^2$.

The presence of a polarized proton target with a high degree of polarization
motivates the study of  polarization effects
(including double spin correlations) in the processes such as $e \vec p \to e \vec p$, $e \vec p \to \vec e p$,
$\vec e \vec p \to e p$ in order to find possibilities for setting up polarization experiments
of a new type for measuring the elastic proton form factors.

In Refs.~\cite{JETPL2008,JETPL18,JETPL19,JETPL2021,PEPAN2022,JETPL2022},
in the OPE approximation,  polarization effects in the elastic
$e \vec p \to e \vec p$ process were investigated in the case when the spins of the initial
and of the detected recoil proton are parallel, i.e., when an proton is scattered in
the direction of the spin quantization axis of the rest proton target.
To do this, in the kinematics of the SANE Collaboration experiment \cite{Liyanage2020}
on measuring double-spin asymmetry in the $\vec e\vec p \to e p$ process, using the Kelly
\cite{Kelly2004} and Qattan \cite{Qattan2015} parametrizations for the $R$ ratio, a numerical analysis was
carried out of the dependence of the longitudinal polarization degree of the scattered proton
on the square of the momentum transferred to the proton as well as on the scattering
angle of the electron and proton.
In this case, a noticeable sensitivity of the transferred to the proton polarization
to the type of dependence of the ratio $R$ on $Q^2$  was established, and it was also
shown that the violation of the scaling of the SFFs leads to a significant increase in
the magnitude of the polarization transfer to the proton, as compared to the case
of the dipole dependence.
Thus, in Refs. \cite{JETPL2008,JETPL18,JETPL19,JETPL2021,PEPAN2022,JETPL2022},
the 4th method for measuring the ratio of $R$ was proposed, 
based on the transfer of polarization from the initial proton to the final
one in the $e \vec p \to e \vec p$ process in the case when their spins are parallel.
This method also works in the TPE approximation and allows
us to measure the squares of the modules of generalized SFFs \cite{JETPL19}.

Note that Akhiezer and Rekalo (see \cite{AR}, pp. 211--215) also performed a  general
calculation of the $e \vec p \to e \vec p$ cross section in the Breit system for
 partially polarized initial and final protons. However, they analyzed this
cross section in \cite{AR} by analogy with \cite{Rekalo74} and overlooked a more interesting case,
which was discussed in Refs. \cite{JETPL2008,JETPL18,JETPL19,JETPL2021,PEPAN2022,JETPL2022}.

In our recent short paper \cite{PRD2023}, the 5th method of measuring the ratio $R$
was proposed, based on the transfer of polarization from the initial proton to the final
electron in the elastic $e\vec p\to \vec e p$ process in the case when the spins quantization
axes of the resting proton target and the scattered electron are parallel, i.e., when
the electron is scattered in the direction of the spin quantization axes of the resting proton target.

In Refs.~\cite{JETPL2021,PEPAN2022,JETPL2022,PRD2023} we utilize two commonly used
parametrization of the SFFs. First one, Kelly parametri\-zation \cite{Kelly2004} (and similar
ones \cite{Brash1,Graczyk2,Sufian3}) based on  a mix of cross section and polarization
data, but without the TPE corrections.  Next one, Qattan parametrization \cite{Qattan2015}
(see also  \cite{Alberico6}) include phenomenological TPE corrections extracted from
the difference between Rosenbluth and polarization measurements. Although there exist
more modern and sophisticated fits to proton data \cite{Borah7,Arrington8}, we restricted
our calculation to the first ones as  it  shows negligible differences in our results  from
different types of parametrizations.

The aim of this article is to give a more detailed view of the results of the work \cite{PRD2023}, as well as
to investigate the double spin asymmetry in the $\vec e \vec p \to ep$ process in the case of parallel
spins of the initial electron and proton.
To do this, in the kinematics of the SANE Collaboration experiment \cite{Liyanage2020}
using the Kelly \cite{Kelly2004} and Qattan \cite{Qattan2015} parametrizations for the SFFs
ratio $R$, a numerical analysis was carried out of the dependence of the longitudinal polarization
degree transferred to the scattered electron in the $e \vec p \to \vec e p$ process
and the double spin asymmetry in the $\vec e \vec p \to e p$ process on the square of the momentum
transferred to the proton, as well as on the scattering angle of the electron.


\section{Helicity and diagonal spin bases}

The spin 4-vector $s=(s_{0}, \vecc s)$ of the fermion with 4-momentum $p$ ($p^2=m^2)$ satisfying
the conditions of orthogonality and normalization is given by
\ba
s=(s_{0}, \vecc s), \; s_{0}=\frac {\vecc c \,\vecc p}{m}, \;
\vecc s =\vecc c + \frac{\,(\vecc c \,\vecc p)\,\vecc p} {m(p_0 +m )\,},
\label{spinq}
\ea
where $\vecc c$ is the spin quantization axis ($\vecc c^{2}=1$).

Expressions (\ref{spinq}) allow us to determine the spin 4-vector $s=(s_{0}, \vecc s)$
by a given 4-momentum $p=(p_{0}, \vecc p)$ and 3-vector $\vecc c$. On the contrary, if the 4-vector
$s$ is known, then the spin quantization axis $\vecc c$ is given by
\ba
\vecc c = \vecc s - \frac{s_0}{p_0+m} \, \vecc p,
\label{spinc}
\ea
i.e. the vectors $\vecc c$ and $s$ at a given $p$ uniquely define each other.

For calculation of  polarization effects in high-energy physics process one usually utilize  helicity basis
introduced by Jacob and Wick \cite{Jacob}, in which the spin quantization axis $\vecc c$
is directed along the momentum of the particle
\ba
\vecc c =\vecc n = \vecc {p}/|\vecc p|, \nn
\ea
while the spin 4-vector $s$ (\ref{spinq}) reads
\ba
s=(s_{0}, \vecc s) =(|\vecc v |, v_{0} \, \vecc n), \nn
\label{spins}
\ea
where $v_{0}$ and $\vecc v$ are the time and space components of the 4-velocity vector $v=p/m$ ($v^2=1$).

The popularity of the helicity basis is primarily due to the simplicity of the physical
interpretation of the helicity definition (projection of the spin in  the direction of the particle momentum),
and  its emphasis on the center of mass system. At the same time,  studying
of helicities of moving particles is analogous to the study of the spins of particles at rest \cite{GL,FIF70}.
However, there are several important factors which prevent helicity from playing the dominant
role in describing the spin projection of particles. One is that  helicity is not a particle
characteristic that is invariant under the Lorentz transformation \cite{GL,FIF70,AB,BLP}.
In interpreting the dynamics of  spin interaction, the amplitudes of scattering processes
with and without changing the sign of the particle helicity are often referred to as
amplitudes with and without a spin flip. However, since the particle momentum
is changed by the interaction, it is clear that such a classification is
very arbitrary. Both types of amplitudes actually describe a process with
a change in the particle spin state.

In general, for a system of two particles with different 4-momenta
$q_1=(q_{10}, \vecc{q_1})$ (before interaction) and $q_2=(q_{20},\vecc{q_2})$
(after interaction) the possibility of quantization of spins in one common direction,
including the case when particles have different masses, is determined
by the three-dimensional vector given by \cite{FIF70}
\ba
\vecc a = \vecc q_{2}/q_{20} - \vecc q_{1}/q_{10}.
\label{osq1q2}
\ea
Since the common spin quantization axis (\ref{osq1q2}) defines the spin basis
other than the helicity one 
and is the difference of two three-dimensional vectors,
the geometric image of which is the diagonal of a parallelogram, it is natural to call
it the diagonal spin basis (DSB). For the first time, in a four-dimensional covariant form,
the DSB was constructed in the work \cite{Sik84} in  the process
\ba
\label{EPEPp1p2}
e(p_1)+ p\,(q_1,s_{p_{1}}) \to e(p_2)+ p \,(q_2,s_{p_{2}}).
\ea
In it, the spin 4-vectors of the initial and final protons $s_{p_{1}}$ and $s_{p_{2}}$
are expressed in terms of their 4-momenta $q_{1}$ and $q_{2}$ ($q_1^2=q_2^2=m^2$) \cite{Sik84}:
\begin{align}
\label{DSBp1p2}
s_{p_{1}} =  \frac { m^2 q_{2} - (q_{1} q_{2}) \,q_{1} } {m \sqrt{(q_{1}q_{2} )^{2} - m^4 }} \, ,\\
%
s_{p_{2}} =   \frac { (q_{1} q_{2})\, q_{2} - m^2 q_{1} } {m \sqrt{(q_{1}q_{2} )^{2} - m^4 }} \,.\nn
\end{align}

In the laboratory frame (LF), where the initial proton rests,
$q_1=(m,\vecc 0)$, the spin 4-vectors (\ref{DSBp1p2}) 
read
\ba
\label{DSB_LSOq1q2}
s_{p_{1}}=(0,\vecc n_2 )\,, \; s_{p_{2}}= (|\vecc v_2|, v_{20}\, \vecc {n_2}),
\ea
where $\vecc n_2=  \vecc {q_2}/|\vecc q_2|$, $v_2=(v_{20}, \vecc v_2)=q_2/m$
is the velocity vector of the final proton,  $v_2^2=1$.

Using the explicit form of the spin 4-vectors
(\ref{DSB_LSOq1q2}) and formulas (\ref{spinc}) or (\ref{osq1q2}), it is easy to verify that
the spin quantization axes of the initial and final proton in the LF
coincide with the direction of the final proton momentum
\ba
\vecc c = \vecc c_{p_{1}} =\vecc c_{p_{2}}=\vecc n_{2}=  \vecc {q_2}/|\vecc q_2|.
\label{LSO}
\ea

In the ultrarelativistic limit, when the masses of protons can be neglected,
i.e. at $q_{10}, q_{20}\gg m$, the spin 4-vectors $s_{p_{1}}$
and $s_{p_{2}}$ (\ref{DSBp1p2}) read
\ba
s_{p_{1}} =  -\frac {q_1}{ m} , \; \; s_{p_{2}} =  \frac {q_2}{ m} \, .
\label{DSBq1q2m0}
\ea

Let us turn to the consideration of the electron-proton scattering $ e \vec p \to \vec e  p$ process
in the case when the initial proton and the final electron are polarized
\ba
e(p_1)+ p\,(q_1,s_{p_{1}}) \to e(p_2,s_{e_{2}})+ p \,(q_2),
\label{EPEPp1e2}
\ea
where $p_1, p_2$ are the 4-momenta of the initial and final electrons ($p_1^2=p_2^2=m_0^2$).

For the process under consideration (\ref{EPEPp1e2}), we define the  common spin quantization axis $\vecc a$
and the spin 4-vectors of the initial proton and the final electron $s_{p_{1}}$ and
$s_{e_{2}}$ as follows:
\begin{align}
\label{osp1e2}
\vecc a & = \vecc p_{2}/p_{20} - \vecc q_{1}/q_{10}, \\
\label{DSBpr1}
s_{p_{1}} &=\frac {m^2 p_{2} - ( q_{1} p_{2})q_{1}}{m\sqrt{ ( q_{1}p_{2} )^{2} - m^2 m_0^2 }}\,, \\
%
s_{e_{2}} &= \frac {( q_{1} p_{2}) p_{2} - m_0^2 q_{1} }{m_0\sqrt{ ( q_{1}p_{2} )^{2} - m^2 m_0^2 }}\,. \nn
\end{align}

In the LF, the  spin 4-vectors (\ref{DSBpr1}) 
read
\ba
\label{DSB_LSO_q1p2}
s_{p_{1}}=(0,\vecc n_{e_{2}} ), \; s_{e_{2}}= (|\vecc {v_{e_{2}}}|, v_{e_{20}}\, \vecc {n_{e_{2}}}),
\ea
where  $\vecc n_{e_{2}}=  \vecc {p_2}/|\vecc p_2|$,   $v_{e_{2}}=(v_{e_{20}}, \vecc {v_{e_{2}}})=p_2/m_0$
is the velocity of the final electron,  $v_{e_{2}}^2=1$.

Using the explicit form of the spin 4-vectors (\ref{DSB_LSO_q1p2}) and formulas (\ref{spinc})
or (\ref{osp1e2}), it is easy to verify that the spin quantization axes of the initial proton
$\vecc c_{p_{1}}$ and the final electron $\vecc c_{e_{2}}$ in the LF coincide with
the direction of the final electron momentum
\ba
\vecc c = \vecc c_{p_{1}} =\vecc c_{e_{2}}=\vecc n_{e_{2}}=  \vecc {p_2}/|\vecc p_2|.
\label{osq1p2}
\ea

In the ultrarelativistic limit, when the electron mass  can be neglected,
i.e. at $p_{10}, p_{20}\gg m_0$, the spin 4-vectors (\ref{DSBpr1}) 
read
\ba
s_{p_{1}} =  \frac { m^2 p_{2} - ( q_{1} p_{2})\,q_{1} }
{m ( q_{1}p_{2} )}, \; \; s_{e_{2}} =  \frac {p_2}{ m_0} \, .
\label{DSBq1p2m0}
\ea

Similarly, in the case when the initial electron and proton are polarized in the $ep$ scattering
process
\ba
e(p_1,s_{e_{1}}) + p (q_1,s_{p_{1}}) \to e(p_{2}) + p(q_2) 
\label{EPEPe1p1}
\ea
the  common spin quantization axis $\vecc a$ and spin 4-vectors of the initial electron
and the proton $s_{e_{1}}$ and $s_{p_{1}}$ are defined as follows:
\begin{align}
\label{osp1q1}
\vecc a & = \vecc p_{1}/p_{10} - \vecc q_{1}/q_{10} ,\\
\label{DSBe1p1}
s_{e_{1}} &=\frac { ( q_{1} p_{1})\,p_{1} - m_0^2\, q_{1}  } {m_0\sqrt{ ( q_{1}p_{1} )^{2} - m^2 m_0^2 }}\, , \\
%
s_{p_{1}} &= \frac { m^2 p_{1} - ( q_{1} p_{1})\,q_{1} } {m\sqrt{ ( q_{1}p_{1} )^{2} - m^2 m_0^2 }}\, \nn.
\end{align}

Again, in the LF, the  spin 4-vectors (\ref{DSBe1p1}) 
read
\ba
\label{DSB_LSO_q1p1}
s_{p_{1}}=(0,\vecc n_{e_{1}} ),  \; s_{e_{1}}= (|\vecc {v_{e_{1}}}|, v_{e_{10}} \vecc {n_{e_{1}}}),
\ea
where $\vecc n_{e_{1}}=  \vecc {p_1}/|\vecc p_1|$,   $v_{e_{1}}=(v_{e_{10}},
\vecc {v_{e_{1}}})=p_1/m_0$  is the velocity of the initial electron, $v_{e_{1}}^2=1$.

Using the explicit form of the spin 4-vectors (\ref{DSB_LSO_q1p1}) and formulas (\ref{spinc})
or (\ref{osp1q1}), it is easy to verify that the spin quantization axes of the initial proton $\vecc c_{p_{1}}$
and  electron  $\vecc c_{e_{1}}$ in the LF coincide with the direction of the initial electron momentum
\ba
\vecc c = \vecc c_{e_{1}}= \vecc c_{p_{1}} =\vecc n_{e_{1}}=  \vecc {p_1}/|\vecc p_1|.
\label{osq1p1}
\ea

In the ultrarelativistic limit, when the electron mass  can be neglected,
i.e. at $p_{10}, p_{20}\gg m_0$, the spin 4-vectors (\ref{DSBe1p1}) 
read
\ba
s_{e_{1}} =  \frac {p_1}{ m_0} , \; s_{p_{1}} =  \frac { m^2 p_{1} - ( q_{1} p_{1})\,q_{1} }
{m ( q_{1}p_{1} )}\,.
\label{DSBq1p1m0}
\ea

Thus, in this section, three DSBs  corresponding to the
$e \vec p \to e \vec p$, $e \vec p \to \vec e p$, $\vec e \vec p \to e p$ processes are built,
of which the last two are considered here for the first time.

The fundamental fact that the Lorentz little group common to a system of two
particles with different momenta is realized in the DSB leads to a number
of remarkable consequences. 
First, in this basis, particles before and after interaction in the scattering channel
have common spin operators \cite{Sik84,GS98}, which allows one to covariantly separate
interactions with and without changing of the spin states of the particles involved in the reaction,
making it possible to trace the dynamics of the spin interaction.
Second, in the DSB, the mathematical structure of the amplitudes is maximally
simplified owing to the coincidence of the particle spin operators, the separation of Wigner
rotations from the amplitudes \cite{Sik84,GS98}, and the decrease in the number of independent
scalar products of 4-vectors that characterize the reaction. Third, in the DSB,
the spin states of massless particles coincide up to the sign with the helicity states,
see Eqs. (\ref{DSBq1q2m0}).

\section{Kinematics and variables used}

The differential cross sections of the processes (\ref{EPEPp1p2}), (\ref{EPEPp1e2}) and (\ref{EPEPe1p1})
calculated in the DSB can in principle  contain only dot products
of the particles 4-momenta $p_i p_j$, $p_iq_j$, $q_iq_j$ $(i,j=1,2)$ involved in reactions.
A further significant simplification of expressions can be achieved by moving
from the 4-vectors $p_i$, $q_i$ to the 4-vectors $p_{\pm}=p_2 \pm p_1, \, q_{\pm}=q_2 \pm q_1$.
They satisfy the orthogonal conditions $p_{\pm}p_{\mp}=q_{\pm}q_{\mp}=p_{\pm}q_{\mp}=0$
and the following simple relations:
\ba
\label{ppqm}
p_+^2 + p_-^2=4 m_0^2, \; q_+^2 + q_-^2=4 m^2, \nonumber \\
 q_+^2=4m^2(1+\tau_p), \; \tau_p=-q_-^2/4m^2.
\ea
In terms of $p_{\pm}$, $q_{\pm}$ the  4-vectors of $p_i$, $q_j$ are: 
\ba
&&p_1=(p_+ -p_-)/2, \, p_2=(p_+ + p_-)/2, \nn \\
&&q_1=(q_+ -q_-)/2, \; q_2=(q_+ + q_-)/2.  \nn
\ea

Let us introduce the orthonormal vector basis (tetrad) $b_{A}$ $(A = 0, 1, 2, 3)$: 
\begin{align}
\label{OBVp}
b_0&=\frac{q_+}{\sqrt{q_+^2}\,}\,, \; b_{3} = \frac{q_-}{\sqrt{-q_- ^2}\,}\,, \\
(b_{2})_{\mu}& = \varepsilon_{\mu \nu \kappa \sigma}q_1^{\nu}q_2^{\,\kappa}r^{\sigma}/\rho,\;
(b_1)_{ \mu} = \varepsilon_{\mu \nu \kappa \sigma}b_0^{\nu}b_3^{\kappa}b_2^{\sigma}, \nn
\end{align}
where $\varepsilon_{\mu\nu\kappa\sigma}$ is the Levi-Civita tensor
($\varepsilon_{1230}=$1); $r$ is the 4-momentum of the particle involved in the reaction which is
different from $q_{1}$ and $q_{2}$; $\rho$ is determined from the normalization conditions
\ba
&& b_{0}^{2}=- b_{1}^{2} = -b_{2}^{2} = -b_{3}^{2}=1. \nn
\ea
The completeness relation is valid for the tetrad of the 4-vectors $b_{A}$ (\ref{OBVp})
\ba
{b_0}_{\mu} \, {b_0}_{\nu} - {b_1}_{\mu} \, {b_1}_{\nu} - {b_2}_{\mu}\, {b_2}_{\nu}
- {b_3}_{\mu} \, {b_3}_{\nu}=g_{\mu \nu} ,
\label{gmunu}
\ea
where $g_{\mu\nu}$ is the metric tensor in the Minkowski space,
which is naturally divided into the sum of the longitudinal and transverse parts:
\ba
&&g_{\mu \nu}=g_{\mu \nu}^{\|}+g_{\mu \nu}^{\bot}, \nn\\
\label{g_parallel}
&&g_{\mu \nu}^{\|}={b_0}_{\mu} \, {b_0}_{\nu}-{b_3}_{\mu} \, {b_3}_{\nu} ,\nn\\
&&g_{\mu \nu}^{\bot}=- {b_1}_{\mu} \, {b_1}_{\nu} - {b_2}_{\mu}\, {b_2}_{\nu}. \nn
\ea
For the transverse part of the metric tensor we have
\ba
g_{\mu \nu}^{\bot}=g_{\mu \nu} - g_{\mu \nu}^{\|}. \nn
\ea
In terms of $(q_{\pm})_{\mu}$, $(q_{\pm})_{\nu}$, the tensor $g^{\bot}_{\mu\nu}$ has the form
\ba
\label{gperp}
g^{\bot}_{\mu \nu}=g_{\mu \nu}-\frac{(q_+)_{\mu} (q_+)_{\nu}}{q_+^2}
+\frac{(q_-)_{\mu} (q_-)_{\nu}\,}{-q_-^2}.
\ea
For calculations we also use the  Mandelstam variables
\ba
s=(p_1+q_1)^2, \, t=(q_2-q_1)^2, \,u=(q_2-p_1)^2,
\label{stu}
\ea
with the  standard connection equation
\ba
s+t+u=2m_0^2 + 2m^2. \nn
\ea
By reversing the relation in  Eqs. (\ref{stu}), for scalar products in terms of $s, t, u$ we have:
\ba
&& 2p_1q_1=2p_2q_2=s-m_0^2-m^2,\; \nn \\
\label{scalprod}
&& 2p_1q_2=2p_2q_1=m_0^2+m^2-u,\, \\
&&2p_1p_2=2m_0^2-t, \;p_+q_+=s-u, \nn \\
&&2q_1q_2=2m^2-t,  \, q_+^2=4m^2-t. \nn
\ea

\subsection{Ultrarelativistic limit}
In the ultrarelativistic limit, when the mass of an electron can be neglected,
for the Mandelstam variables in the LF we have:
\begin{align}
 s&=m^2+2m E_1, \nn\\
 -t &= Q^2 = 4E_1 E_2\sin^2 (\theta_e/2), \nn \\
  u&=m^2-2 m E_2, \nn
\end{align}
where the $\theta_e$ is the angle between the vectors $\vecc p_1$ and $\vecc p_2$,
$\cos(\theta_e)=\vecc p_1 \vecc p_2/|\vecc p_1| |\vecc p_2|$.

The energies of the final electron $E_2$ and the proton $E_{2p}$ are related in the LF
with $Q^2=-q_-^2$ as follows:
\ba
\label{E2Q}
&&E_2=E_1-Q^2/2m, \;  E_{2p}=m+Q^2/2m.   
\ea

For the dot products $q_+^2, p_+q_+$, and $q_-^2$  we have:
\ba
\label{skalProd}
&&q_+^2=4m^2+2m E_-,\;   p_+q_+=s-u=2m E_+,   \\
&&q_-^2=-2m E_-, \; E_{2p}=m+E_-, \, E_{\pm}=E_1 \pm E_2. \nn
\ea

The dependences of $E_2$ and $Q^2$ on the electron scattering angle $\theta_e$ in the LF are
\ba
\label{E2tea}
 E_2(\theta_e)&=&\frac{E_1}{1+ (2E_1/m)\, \sin^2 (\theta_e/2)}, \\
\label{Q^2te}
 Q^2(\theta_e)&=&\frac {4 E_1^{\,2} \sin^2 (\theta_e/2)}{1+ (2E_1/m) \sin^2 (\theta_e/2)}.
\ea

The dependence of $E_{2p}$ and $Q^2$ on the proton scattering angle $\theta_p$ has the form
\ba
\label{E2tp}
E_{2p}(\theta_p)&=&m\, \frac{(E_1+m)^2+E_1^{\,2}\cos^2(\theta_p)}
{(E_1+m)^2-E_1^{\,2}\cos^2(\theta_p)}\,,\\
\label{Q^2tp}
Q^2(\theta_p)&=&\frac{4m^2 E_1^{\,2}\cos^2(\theta_p)}
{(E_1+m)^2-E_1^{\,2}\cos^2(\theta_p)}\,,
\ea
where the $\theta_p$ is the angle between the vectors $\vecc p_1$ and $\vecc q_2$,
$\cos(\theta_p)=\vecc p_1 \vecc q_2/|\vecc p_1| |\vecc q_2|$.

The inverse relations between $\theta_e$, $\theta_p$ and $E_2$, $E_{2p}$ can be written as
\ba
&&\label{tetae}
\theta_e=\arccos{\left(1-\frac{Q^2}{2 E_1 E_2}  \right)}, \\
\label{tetap}
&& \theta_p=\arccos{\left(\frac{E_1+m}{E_1} \sqrt{\frac{\tau_p}{1+\tau_p}}\;\right)}.
\ea

In the elastic $e p \to e p $ process the electron scattering angle $\theta_e$
changes from $0^{\circ}$ to $180^{\circ}$, while $Q^2$ changes in the range
of $0 \leqslant Q^2 \, \leqslant Q^2_{max}$ ($0 \leqslant \tau_p \, \leqslant \tau_{max}$), where
\ba
\label{Qmax}
Q^2_{max}=\frac{4ME_1^{\,2}}{(M+2E_1)}, \,\,
\tau_{max}=\frac{E_1^{\,2}}{M(M+2E_1)}.
\ea
Let us write the following useful relation:
\ba
\label{taum+1}
\sqrt{\frac{\tau_{max}}{1+\tau_{max}}}=\frac{E_1}{M+E_1}.
\ea

According to Eq. (\ref{Q^2te}), at $\theta_e=0$ we have $Q^2=0$
and $\tau_p=0$. However, from  Eq. (\ref{tetap}) it follows that 
in this case $\theta_p=90^{\circ}$.
In the case of electron backscattering ($\theta_e=180^{\circ}$),
when $\tau_p=\tau_{max}$, it follows from Eqs. (\ref{tetap}) and (\ref{taum+1})
that $\theta_p=0^{\circ}$. Thus, the electron scattering
by an angle ranging from $0^{\circ}$ to $180^{\circ}$
($0^{\circ} \leqslant \theta_e \, \leqslant 180^{\circ}$)
leads to a change in the proton scattering angle from
$90^{\circ}$ to $0^{\circ}$.

\begin{figure}[h!]
\centering
\includegraphics[width=0.50\textwidth]{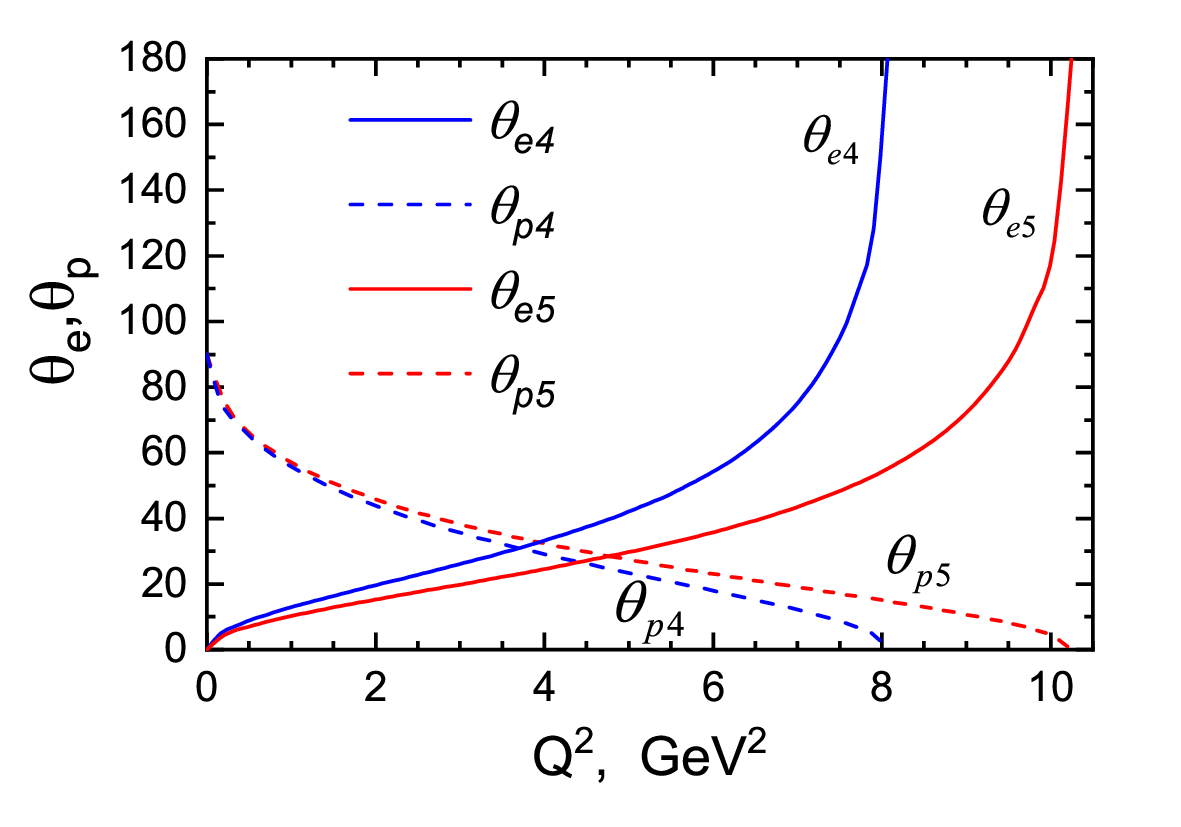}
\vspace{-8mm}
\caption{  
$Q^2$ dependence of the scattering angles of the electron $\theta_e$ and the proton $\theta_p$
(in degrees) at electron beam energies in the experiment \cite{Liyanage2020}. The lines $\theta_{e4}$,
$\theta_{p4}$ ($\theta_{e5}$, $\theta_{p5}$) correspond to $E_1=4.725$ ($5.895$)~$\GEV$.
}
\label{Theta_ep1}
\end{figure}

The results of calculations of the dependence of the scattering angles of the electron $\theta_e$
and proton $\theta_p$ on the square of the momentum transferred to the proton $Q^2$ at electron
beam energies $E_1=4.725$ GeV and $5.895$ GeV in the SANE Collaboration experiment \cite{Liyanage2020}
are plotted in Fig. \ref{Theta_ep1}. They correspond to the lines labeled $\theta_{e4},
\theta_{p4}$ and $\theta_{e5}, \theta_{p5}$.

\vspace{-3mm}

The intersection points of the lines $\theta_{e4}$ and $\theta_{p4}$ ($\theta_{e5}$ and $\theta_{p5}$)
in Fig. \ref{Theta_ep1} correspond to the equality $\theta_{e}=\theta_{p}$
for some values $Q^{\,2}=Q_{(ep)}^{\,2}$. At the same time
$Q^{\,2}_{(ep)}=3.70\,\GEV^{2}$ for $E_1=4.725$ GeV and $Q^{\,2}_{(ep)}=4.772\, \GEV^{2}$
for $E_1=5.895$ GeV. For the corresponding angles we have
$\theta_{ep}=30.91^{\circ}$ (0.54 rad) and $\theta_{ep}=28.45^{\circ}$ (0.50 rad).

The data on the electron and proton scattering angles (in radians) at electron beam energies
$E_1=4.725$ GeV and 5.895 GeV and $Q^2=2.06$ GeV$^2$ and 5.66 GeV$^2$ are represented in Table \ref{Uglyep},
which contains also the values of $Q^2_{max}$ (\ref{Qmax}) for the maximally possible $Q^2$
values at $E_1=4.725$ GeV and 5.895 GeV.

\begin{table}[h!]
\centering
\caption{
Electron and proton scattering angles $\theta_e$ and $\theta_p$ (in radians) in the
kinematics of the  experiment \cite{Liyanage2020}.
}
\label{Uglyep}
\tabcolsep=1.5mm
\footnotesize
\begin{tabular}
{| c | c | c | c | c | c | }
\hline
$E_1$ (\rm{GeV})
&  $Q^2$ (\rm{GeV}$^2$)
& $\theta_{e}\, (rad)$ & $\theta_{p}\, (rad)$
& $Q^2_{max}$ (\rm{GeV})$^2$   \\
\hline
5.895 &2.06 & 0.27 &  0.79  & 10.247  \\
\hline
5.895 &5.66 & 0.59 & 0.43  & 10.247 \\
\hline
4.725 &2.06 & 0.35 & 0.76  & 8.066  \\
\hline
4.725 &5.66 & 0.86 & 0.35 & 8.066 \\
\hline
\end{tabular}
\end{table}

\section{Polarization of a virtual photon in the $e p \to e p$ process}

The $\varepsilon$ value entering into the expression for the Rosenbluth cross section
(\ref{Ros}) with the range of variation $0 \leqslant \varepsilon \leqslant 1$
in modern literature, as a rule, is identified not with the degree of linear
(transverse) but with the degree of longitudinal polarization of the virtual photon.
Sometimes it is also referred to as the polarization parameter or simply the virtual photon
polarization.

 For example,  in  Ref. \cite{Tomalak:2014dja} $\varepsilon$ in the massless case  was interpreted
as a  degree of the longitudinal polarization in the  OPE approximation.
Similar statements have been made in a number of other works
\cite{Jones00,Qattan2005,Arrington:2011dn,Pun05,Puckett12}.

 The  correct understanding of the physical meaning of the value $\varepsilon$
is quite rare \cite{Gakh2008,Alguard1976,Alguard1976b}, but recently the
number of such works has been gradually increased, see, for example,
\cite{Tomalak:2018jak,Korchagin2021,Weiss2023}.

The most common expression in the literature for $\varepsilon$, given on the first page,
actually contains the  dependence on the electron scattering angle $\theta_e$ in the LF.
Expressions for $\varepsilon$, that make it possible to calculate the dependences
of the quantities of interest on, e.g., $Q^2$ or the proton scattering angle $\theta_{p}$
are given by
\begin{align}
\label{eps2}
\varepsilon^{-1}& = 1+\frac{(E_1-E_2)^2+2 (E_1-E_2)\,m}{2 E_1 E_2-(E_1-E_2)\,m}\\
&=\frac{E_1^{\,2}+E_2^{\,2}+(E_1-E_2)\,m} {2 E_1 E_2-(E_1-E_2)\,m}\,,\nn
\end{align}
where $E_1$ and $E_2$ are the energies of the initial and final electrons, respectively.
Note that Eqs. (\ref{E2Q}) 
should  be used for $E_2$, they depend
explicitly only on  $Q^2$; In turn, the $Q^2$ dependence on the angles of $\theta_{e}$
or $\theta_{p}$ is determined by Eqs. (\ref{Q^2te}) or (\ref{Q^2tp}).

\begin{figure}[h!]
\centering
\includegraphics[width=0.50\textwidth]{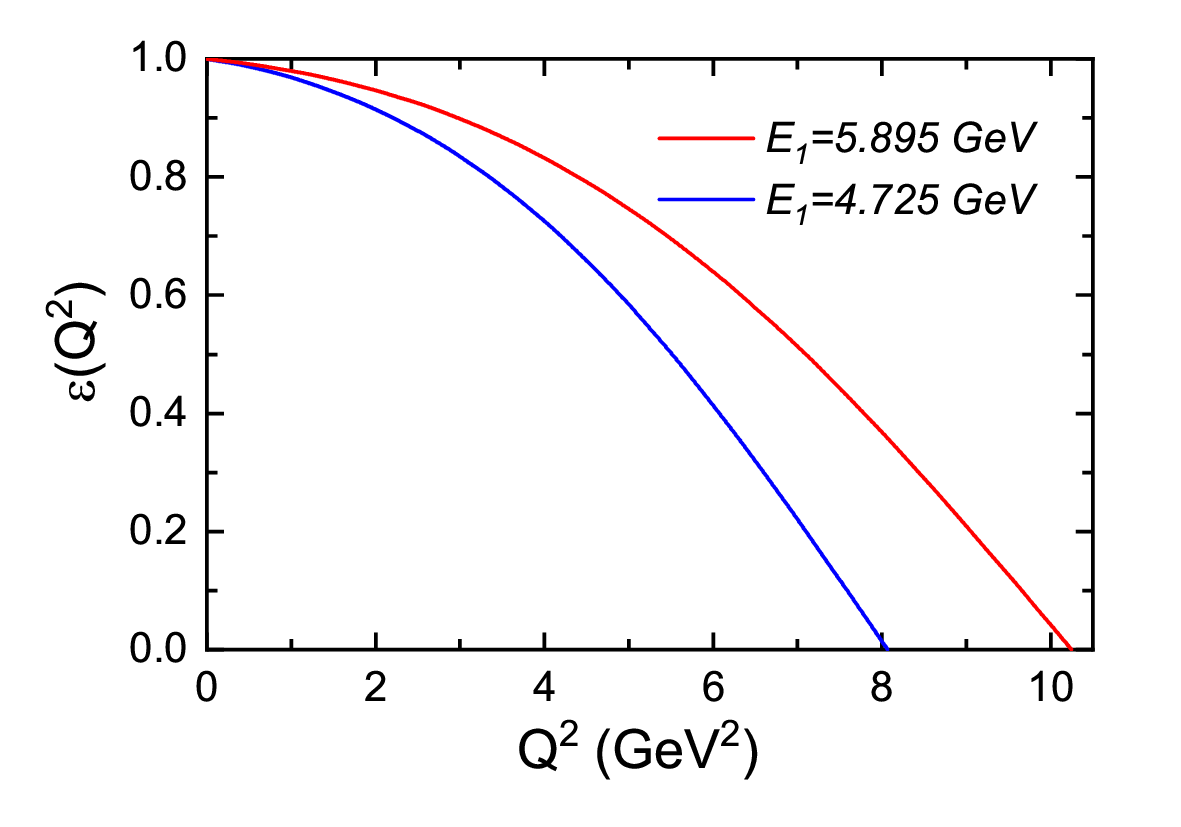}
\vspace{-6mm}
\caption{  
$Q^2$ dependence of the degree of the linear polarization of the virtual photon, $\varepsilon$
(\ref{eps2}), for the electron beam energies 
used in the experiment \cite{Liyanage2020}.
}
\label{eps_01}
\end{figure}

The $Q^2$ dependence of the degree of the linear polarization of the virtual photon,
$\varepsilon$ (\ref{eps2}), at electron beam energies in the SANE Collaboration experiment
\cite{Liyanage2020} is represented by graphs in the Fig.~\ref{eps_01}.

\begin{figure}[h!]
\centering
\includegraphics[width=0.50\textwidth]{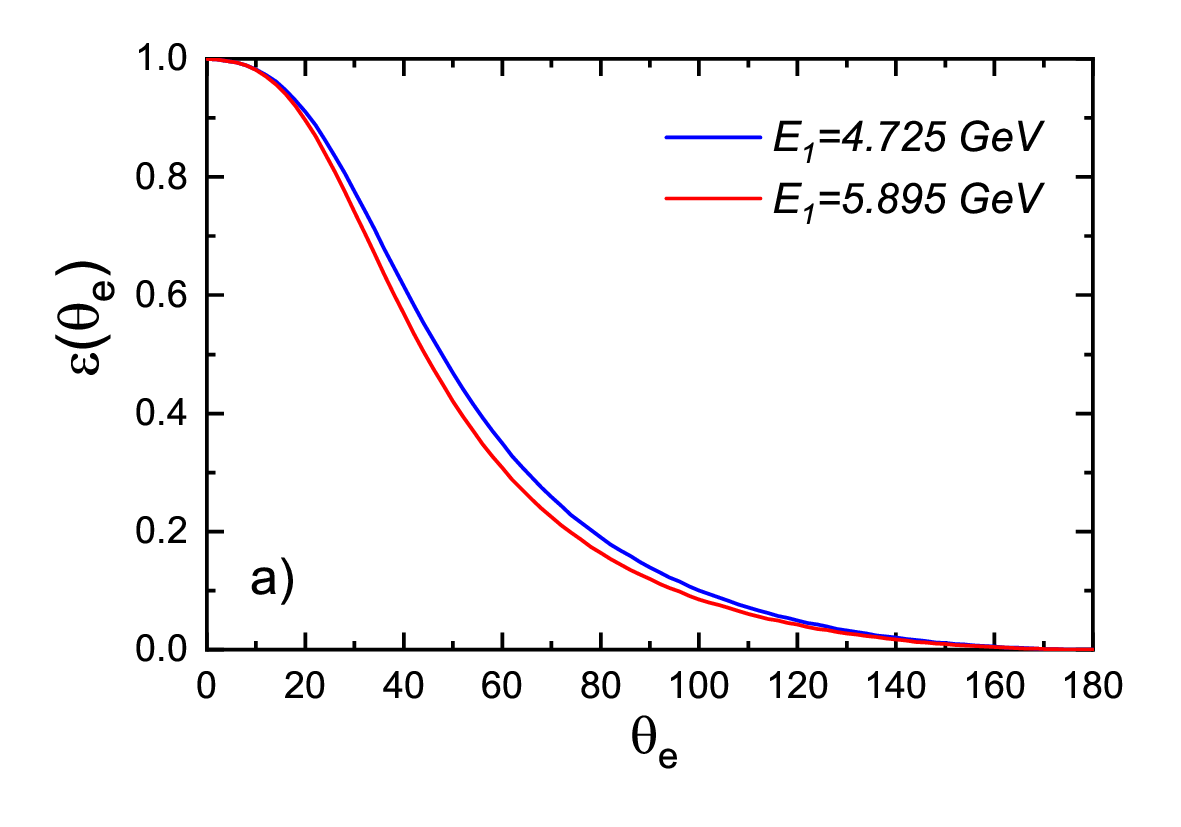}
\vspace{-5mm}
\includegraphics[width=0.50\textwidth]{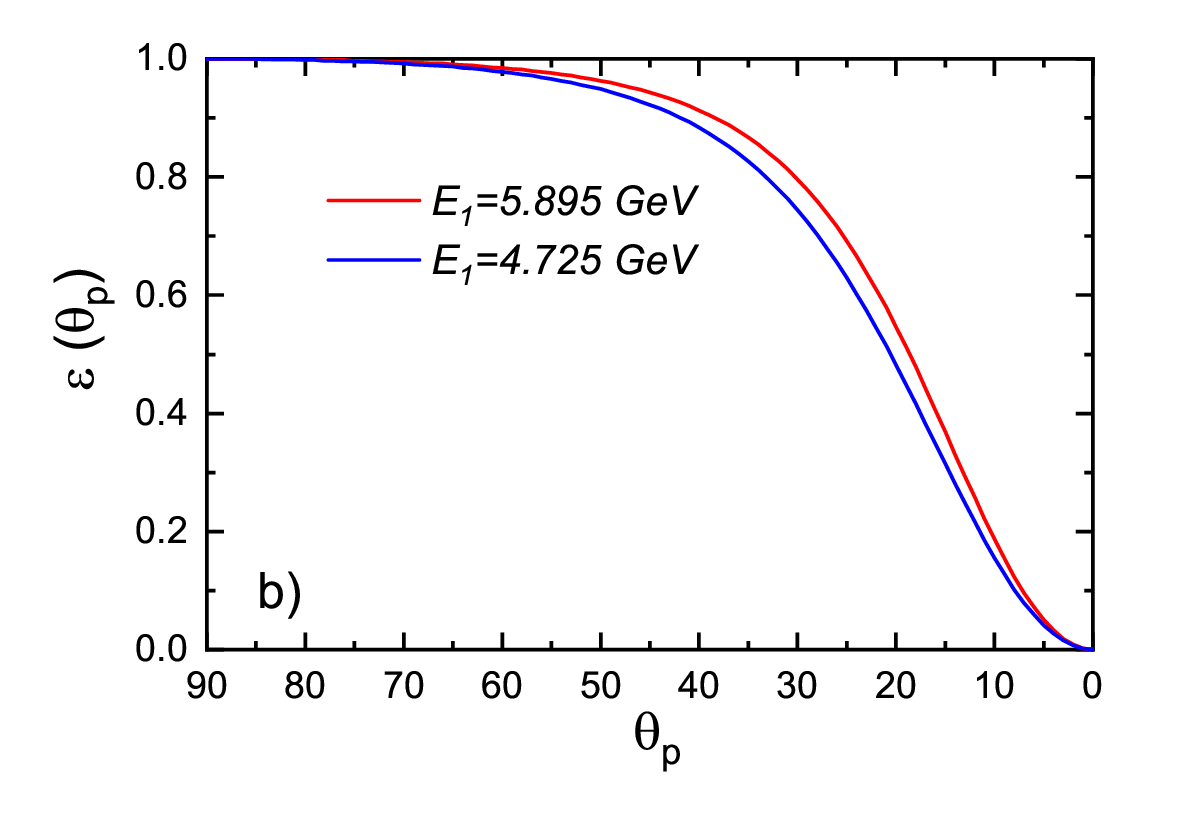}
\caption{ 
Angular dependence of the degree of the linear polarization of the virtual photon, $\varepsilon$
(\ref{eps2}), on the scattering angles of the electron $\theta_e$ {(\bf a)} 
and proton $\theta_p$ {(\bf b)} expressed in degrees for the electron beam
energies in the experiment \cite{Liyanage2020}.
}
\label{epsilon_from_theta_ep}
\end{figure}

Figure \ref{epsilon_from_theta_ep} shows the dependence
of the degree of the linear polarization of the virtual photon, $\varepsilon$ (\ref{eps2}),
on the scattering angles of the electron $\theta_e$ {\bf (a)}
and proton $\theta_p$ {\bf (b)} for the electron beam energies $E_1=4.725$ GeV and $5.895$ GeV
in the experiment \cite{Liyanage2020}.

It follows from Fig.~\ref{eps_01} that $\varepsilon$ is a function of $Q^2$ and decreases
from $1$ to $0$. It follows from Fig.~\ref{epsilon_from_theta_ep}(a) that in the case of electron scattered
forward ($\theta_e=0^\circ$) when $Q^2=0$, $\varepsilon=1$; for a backscattered electron
($\theta_e=180^\circ$) when $Q^2=Q^2_{max}$, $\varepsilon=0$, what is in agreement with
the results of the Ref. \cite{Bernauer2021}.
%

The $Q^2_{max}$ values for
the energies $E_1=4.725$ GeV and $5.895$ GeV are listed in Table \ref{Uglyep}; they amount
to $8.066$ GeV$^2$ and $10.247$ GeV$^2$, respectively. Specifically at these points
the lines in Fig. \ref{eps_01} intersect the abscissa axis.

\section{Polarization effects in the $e \vec p \to \vec e  p$ process}

\subsection{Differential cross section of the $e \vec p \to \vec e  p$ process}

In the OPE approximation, the matrix elements of the process $e p\to  e p$ 
are the product of the electron $(J_{e})$ and proton currents $(J_{p})$
\ba
\label{Mepep}
&& M_{ep\to ep} = 4\pi \alpha \, T_{ep} / q^2,  \\
&& T_{ep} =(J_{e})^{\mu} ( J_{p} )_{\mu} \,.
\label{ampT}
\ea
The lepton $(J_{e})^{\mu}$ and proton $(J_p)_{\mu}$ currents read:
\ba
\label{Je}
 (J_{e})^{\mu} &=& \overline{u}(p_{2}) \gamma^{\mu} u(p_{1}) ,  \nn \\
\label{Jp}
 (J_p)_{\mu} &=&\overline{u}(q_2) \Gamma_{\mu}(q^{2}) u(q_1) , \nn \\
\Gamma_{\mu}(q^{2}) &=& F_{1} \gamma_{\mu} + \frac{F_{2}} {4m}
( \; \hat q \gamma_{\mu} - \gamma_{\mu} \hat q \; ).  \nn
\label{Gamuepep}
\ea
Here $u(p_{i})$ and $u(q_{i})$ are the bispinors of electrons and protons
with the 4-momenta $p_{i}$ and $q_{i}$, respectively, where
$p_{i}^{2} = m_0^{2}$ and $q_{i}^{2} = m^{2}$, having the properties
$\overline{u}(p_{i})u(p_{i})=2m_0$ and $\overline{u}(q_{i}) u(q_{i})= 2m$ $(i = 1,2)$;
$F_{1}$ and $F_{2}$ are the Dirac and Pauli proton form factors, respectively;
$q = q_-=q_{2}-q_{1}$ is the 4-momentum transferred to the proton;
$\hat{q}=(q)_{\mu}\gamma^{\mu}$, where $\gamma^{\mu}$ 
are the Dirac matrices.

It is well known that the SFFs  $G_E$ and $G_M$ could be expressed in terms of
the Dirac and Pauli proton form factors:
\ba
G_{E} = F_{1} -\tau_p F_{2} \, , \; G_{M} = F_{1} + F_{2}. \nn
\label {FFSep}
\ea

The differential cross section of the process $ep \to ep$ 
reads
\ba
\label{Modepep}
\frac {d \sigma_{ep \to ep}}{d t}= \frac{ \pi \alpha^2 }{\lambda_s}\, 
\frac {|T_{ep}|^2}{t^2} \, ,
\ea
where $\lambda_s=(s-(m+m_0)^2)(s-(m-m_0)^2)$ is K{\"a}ll{\'e}n's function.

In the standard approach \cite{AB,BLP}, the calculation of the squares of the amplitude
modules $|T_{ep}|^2$ 
is reduced to the convolution of the lepton ($L^{\mu\nu}$)
and hadron ($H_{\mu\nu}$) tensors
\ba
|T_{ep}|^2=H_{\mu \nu}L^{\mu \nu}, \nn
\ea
where
\ba
\label{Lmunu}
&&L^{\mu \nu}= \Tr(\tau_{e_{2}} \gamma^\mu \tau_{e_{1}} \gamma^\nu),\\
\label{Hmunu}
&& H_{\mu \nu}=2\, 
\Tr(\tau_{p_{2}} \,\Gamma_\mu \tau_{p_{1}} \,\overline{\Gamma}_\nu)\,.~~~
\ea
Here symbol ``\rm{Tr}'' denotes the operation to calculate the trace from Dirac's operators,
$\tau_{e_{i}}$ and $\tau_{p_{i}}$ $(i=1,2)$ are the polarization density matrices
of the initial and final states of electrons and protons ($\lambda_{p_{1}}$ and
$\lambda_{e_{2}}$ are the degrees of polarization of the initial proton
and the final electron, and $\gamma_5$ is the Dirac's matrix):
\ba
&& \tau_{e_{1}}=(\hat{ p_1} +m_0)/2,\, \nn \\
&& \tau_{e_{2}}=(\hat{ p_2} +m_0)(1-\lambda_{e_{2}}\gamma_5 \hat{s}_{e_{2}})/2\,,\\
&& \tau_{p_{1}}=(\hat{ q_1} +m)(1-\lambda_{p_{1}}\gamma_5 \hat{s}_{p_{1}})/2\,,\nn \\
&& \tau_{p_{2}}=(\hat{ q_2} +m)/2\,.\nn
\ea

The lepton tensor $L^{\mu\nu}$ (\ref{Lmunu}) in terms of the 4-vectors $p_{\pm}$ has the form:
\ba
\label{Lmunu1}
2L^{\mu \nu} &=&p_+^{\mu} p_+^{\nu} - p_-^{\mu} p_-^{\nu} + p_-^2 g^{\mu \nu}
+2 i m_0 \lambda_{e_{2}} \varepsilon^{\mu \nu \rho \sigma} (p_-)_{\rho} (s_{e_{2}})_{\sigma}.\nn
\ea

In the ultrarelativistic limit when $s_{e_{2}} = p_2/m_0$, it takes the form
\ba
\label{Lmunu0}
2L^{\mu \nu} &=&p_+^{\mu} p_+^{\nu} - p_-^{\mu} p_-^{\nu} + p_-^2 g^{\mu \nu}
+ i \lambda_{e_{2}} \varepsilon^{\mu \nu \rho \sigma} (p_-)_{\rho} (p_+)_{\sigma}.\nn
\ea
The explicit form of the tensor $H_{\mu\nu}$ (\ref{Hmunu}) is rather cumbersome;
for this reason we omit it.

The expression for $|T_{ep}|^2$ 
can be written as
\ba
2\,|T_{ep}|^2= \frac{4m^2}{q_+^2} |T|^2.  \nn 
\label{T2Y}
\ea
Since $q_+^2=4m^2(1+\tau_p$), then the differential cross section 
of the process (\ref{EPEPp1e2}) calculated in an arbitrary reference frame in the DSB (\ref{DSBpr1}) 
takes the form
\ba
\label{T2all}
\frac {d \sigma_{e \vec p \to \vec e  p }}{d t}&=&\frac{ \pi \alpha^2 }
{2\lambda_s (1+\tau_p)}\,\frac {|{T}|^2}{t^2},\\
\label{T4ya}
|T|^2& =& I_0+\lambda_{p_{1}} \lambda_{e_{2}} I_1, \\
I_0&=&G_E^{\,2} Y_1 + \tau_p\, G_M^{\,2} Y_2, \nn \\
I_1&=& \tau_p (G_E G_M Y_3 + G_M^{\,2} Y_4), \nn
\ea
where $\lambda_{p_{1}}$ and $\lambda_{e_{2}}$ are the degrees of polarization of the initial proton
and the final electron; the functions $Y_i$ ($i=1, \ldots 4$) are given by
\ba
Y_1&=&(p_+q_+)^2+q_+^2q_-^2, \nn \\
\label{Y2}
Y_{2}&=&(p_+ q_+)^2-q_+^2(q_-^2+4 m_0^2),  \\
-Y_{3}&=& 2\,\kappa_1\, m^2\,((p_+q_+)^2+q_+^2(q_-^2-4\,m_0^2\,))\,z_1^2,\nn \\
Y_{4}&=& 2 (m^2 p_+ q_+ -\kappa_1 q_+^2)(\kappa_1 p_+ q_+ - m_0^2 q_+^2)z_1^2, \nn \\
&&z_1=   
( \kappa_1 ^{2} - m^2 m_0^2 )^{-1/2}\,, \;\kappa_1=q_{1}p_{2} .\nn
\ea

In the case of arbitrary spin 4-vectors $s_{p_{1}}$ and $s_{e_{2}}$,
the expressions
for $Y_3$ 
and $Y_4$ 
in the cross section (\ref{T2all}) 
are given by
\ba
\label{Y3a}
Y_{3}&=& 8\,m_0 m \, (s_{p_{1}})^{\mu}
(g^{\bot})_{\mu \nu} (s_{e_{2}})^{\nu} q_+^2 \,,\\
Y_{4}&=& 8\,m_0 m \,(q_+ s_{p_{1}}) (q_+ s_{e_{2}}). \nn
\ea
Note, first, that the polarized part of the cross section (\ref{T2all})
includes the term with $Y_3$ containing the product of the SFFs, $G_E G_M$,
and according to (\ref{Y3a}), is determined by the transverse part of the metric
tensor $(g^{\bot})_{\mu\nu}$ (\ref{gperp});
second, in the cross section of the process $e\vec p\to e\vec p$
in DSB (\ref{DSBp1p2}) 
there is no similar structure, see \cite{JETPL2021,PEPAN2022,JETPL2022}.

In the ultrarelativistic limit, when the mass of an electron can be neglected,
for the functions $Y_i$ (\ref{Y2}) in the LF 
we obtain expressions, that depend only on the energy of the initial and final electrons:
\ba
&&Y_1 =8m^2 (2 E_1E_2 - m E_- ), \nn \\
\label{Y2c}
&&Y_2=8 m^2 (E_1^{\,2} + E_2^{\,2} + m E_-),\\
&&Y_3=-(2m/E_2)\, Y_1 , \nn \\
&&Y_4=8 m^2 E_+ E_- (m-E_2)/E_2.\nn
\ea

\subsection{Polarization of the final electron in the $e \vec p \to \vec e p $ process }

The expression for the square of the amplitude modulus $|T|^2$ (\ref{T4ya}) can be written as
\ba
\label{Tp1e2}
&&|T|^2=I_0+\lambda_{p_{1}} \lambda_{e_{2}} I_1=I_0\,(1+\lambda_{e_{2}}\lambda_{e_{2}}^{f}),
\ea
where 
$\lambda_{e_{2}}^{f}$ 
is the degree of longitudinal polarization
transferred from the initial proton to the final electron in the $e\vec p\to \vec e p$ process:
\ba
\lambda_{e_{2}}^{f}=\lambda_{p_{1}}\frac{I_1}{I_0}=\lambda_{p_{1}}
\frac{\tau_p \, (G_E G_M Y_3 + G_M^{\,2} Y_4)}{G_E^{\,2} Y_1 + \tau_p G_M^{\,2} Y_2}. \nn
\ea
Dividing the numerator and denominator in the last expression by $Y_1 G_M^{\,2}$ and introducing
the experimentally measured ratio $R \equiv \mu_p G_E/G_M$, we get:
\ba
\label{lambdae}
\lambda_{e_{2}}^{f}=
\lambda_{p_{1}}\, \frac{\mu_p \tau_p\, ( (Y_3/Y_1) R+ \mu_p (Y_4/Y_1))}
{ R^{\,2} + \mu_p^2 \,\tau_p\, (Y_2/Y_1) \,}.
\ea
Note that for the ratio $Y_2/Y_1$ in the denominator of  expression (\ref{lambdae})
in the LF, the equality $Y_2/Y_1 = 1/\varepsilon$ is valid, where $\varepsilon$ is the degree
of  linear polarization of the virtual photon (\ref{eps2}).

Inverting relation (\ref{lambdae}), we obtain a quadratic equation with respect to $R$
\ba
\alpha_0 R^2 -\alpha_1 R + \alpha_0 \alpha_3-\alpha_2=0
\label{Rp}
\ea
with the coefficients
\ba
\label{alpha1}
&& \alpha_0=\lambda_{e_{2}}^f / \lambda_{p_{1}}, \; \alpha_1=\tau_p\, \mu_p\,  Y_3/Y_1, \nn \\
&& \alpha_2=\tau_p\, \mu_p^2\, Y_4/Y_1, \; \alpha_3=\tau_p\, \mu_p^2\, Y_2/Y_1.\nn
\ea
Solutions to Eq. (\ref{Rp}) read
\ba
\label{Rsqrt}
R=\frac{\alpha_1 \pm \sqrt{\alpha_1^2-4\alpha_0(\alpha_0\alpha_3-\alpha_2)}}{2\alpha_0}. \nn
\ea
They allow us to extract the ratio $R$ from the results of an experiment to measure the polarization
transferred to the electron $\lambda_{e_{2}}^f$ in the $e\vec p\to \vec e p$ process.

\subsection{Results of numerical calculations of polarization effects }

Formulas (\ref{Y2c}) 
were used to numerically calculate the $Q^2$ dependence
of the longitudinal polarization degree of the scattered electron $\lambda_{e_{2}}^f$ (\ref{lambdae})
as well as the dependence on the scattering angles of the electron and proton at electron beam energies
($E_1=4.725$ GeV and $5.895$ GeV) and the polarization degree of the proton target at rest
($\lambda_{p_{1}}=P_t=0.70$) in the experiment \cite{Liyanage2020} while conserving the scaling
of the SFFs in the case of a dipole dependence $R=R_d$ ($R_d=1$), and in the case of its violation.
In the latter case, the parametrization $R=R_j$
\ba
R_j = (1+0.1430\,Q^2-0.0086\,Q^4+0.0072\,Q^6)^{-1}
\label{Rdj}
\ea
from Ref. \cite{Qattan2015} was used and also the Kelly parametrization ($R=R_k$)
from Ref. \cite{Kelly2004} formulas for which we omit.

\begin{figure}[h!]
\centering
\hspace{-9mm}
\includegraphics[width=0.50\textwidth]{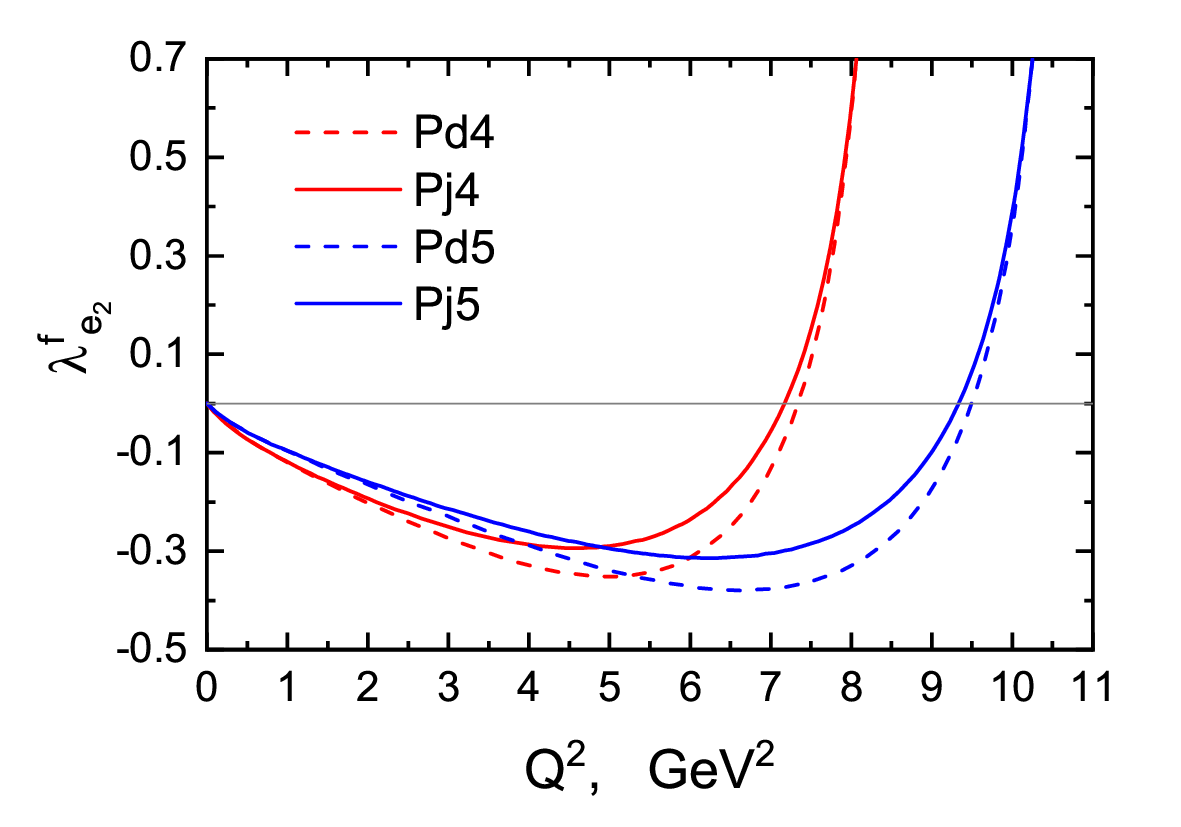}
\vspace{-5mm}
\caption{  
$Q^2$ dependence of the longitudinal polarization degree of the scattered electron
$\lambda_{e_{2}}^{f}$ (\ref{lambdae}) at electron beam energies in the experiment
\cite{Liyanage2020}. The lines $Pd4$, $Pd5$ (dashed) and $Pj4$, $Pj5$ (solid) correspond
to the ratio $R=R_d$ and $R=R_j$ (\ref{Rdj}). The lines $Pd4$, $Pj4$ ($Pd5$, $Pj5$)
correspond to the energies $E_1=4.725$ ($5.895$) $\GEV$.
}
\label{exp12}
\end{figure}

\begin{figure}[h!]
\centering
\includegraphics[width=0.50\textwidth]{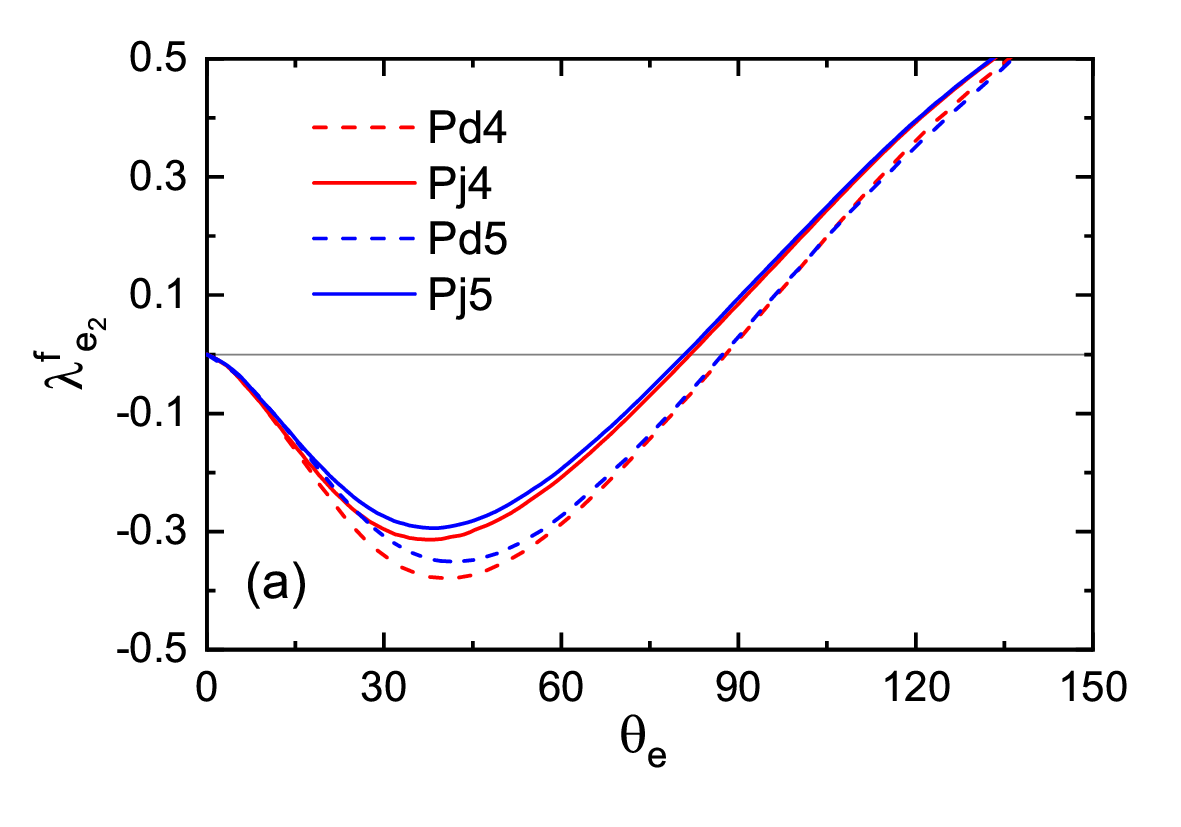} \\
\vspace{-6mm}
\includegraphics[width=0.50\textwidth]{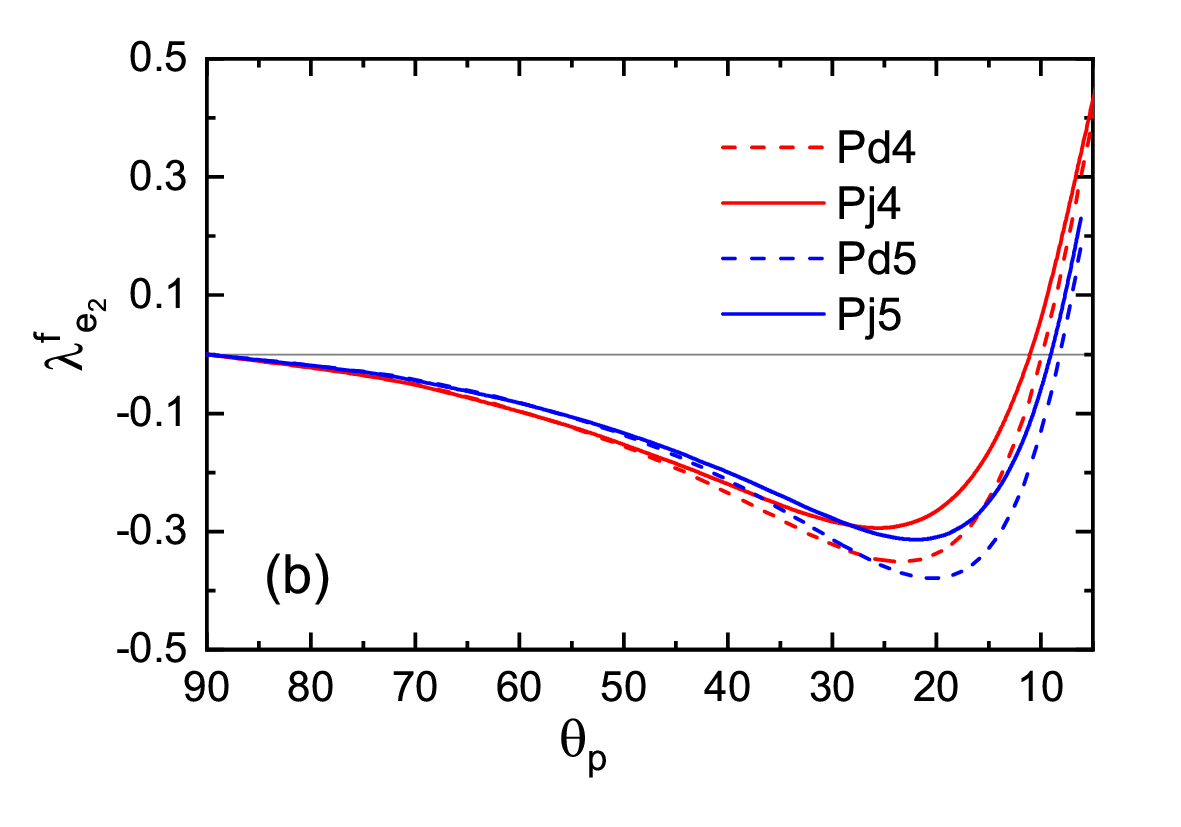} 
\vspace{-9mm}
\caption{ 
Angular dependence of the degree of the transferred to the electron polarization
$\lambda_{e_{2}}^{f}$ (\ref{lambdae}) at electron beam energies used in the experiment
\cite{Liyanage2020} on the scattering angle of the electron $\theta_e$ {\bf (a)}
and of the proton $\theta_p$ {\bf (b)}, expressed in degrees.
The marking of lines $Pd4$, $Pd5$, $Pj4$, $Pj5$ is the same as in Fig.~\ref{exp12}.
}
\label{lambda_from_theta_ep}
\end{figure}

The calculation results are presented by graphs in Figs.~\ref{exp12} and \ref{lambda_from_theta_ep}.
Note that in these figures there are no lines corresponding to the parametrization 
\cite{Kelly2004} since calculations using $R_j$ and $R_k$ give almost identical results.

The $Q^2$ dependence of the longitudinal polarization degree of the scattered electron $\lambda_{e_{2}}^{f}$
(\ref{lambdae}) is plotted in Fig.~\ref{exp12}, on which the lines $Pd4$, $Pd5$ (dashed) and $Pj4$, $Pj5$ (solid)
are constructed for $R=R_d$ and $R=R_j$ (\ref{Rdj}). At the same time, the red lines $Pd4$, $Pj4$
and the blue lines $Pd5$, $Pj5$ correspond to the energy of the electron beam $E_1=4.725$ GeV
and $5.895$ GeV. For all lines in Fig.~\ref{exp12} the degree of polarization of the proton target
at rest $P_t=0.70$.

As can be seen from the graphs in Fig.~\ref{exp12}, the function $\lambda_{e_{2}}^{f}(Q^2)$ (\ref{lambdae})
takes negative values for most of the allowed values $Q^2$ and has a minimum for some of them,
we will specify them: $Pd4(4.976)=-0.352$, $Pj4(4.586)=-0.294$, $Pd5(6.648)=-0.380$, $Pj5(6.254)=-0.314$.
We also give the values for $Q^2$, at which the lines in the Fig.~\ref{exp12} intersect
with the abscissa axis (begin to take positive values): $Pj4(7.174)=0$, $Pd4(7.340)=0$, $5(9.333)=0$,
$ Pd5(9.492)=$0. Thus, in a smaller part of the allowed values adjacent to $Q^2_{max}$ and
amounting to approximately 9\% of $Q^2_{max}$, the function $\lambda_{e_{2}}^{f}(Q^2)$
takes positive values. At the boundary of the spectrum at $Q^2=Q^2_{max}$, the polarization
transferred to the electron is equal to the polarization of the proton target,
$\lambda_{e_{2}}^{f}(Q^2_{max})=P_t=0.70$.

The results of calculations of the angular dependence of the polarization transferred
to the electron $\lambda_{e_{2}}^{f}$ (\ref{lambdae}) in the $e\vec p \to \vec e p$ process
at electron beam energies $E_1=4.725$ GeV and $E_1=5.895$ GeV in the experiment \cite{Liyanage2020}
as functions of the scattering angle of the electron ($\theta_e$) and proton ($\theta_p$)
are represented by graphs in  Fig.~\ref{lambda_from_theta_ep}. The degree of  polarization
of the proton target for all lines was taken to be the same and equal to $P_t=0.70$.
The figures (a) and (b) represent the dependence on the scattering angle of the electron $\theta_e$
and proton $\theta_p$, respectively.

Obviously, the behavior of the lines in Fig.~\ref{lambda_from_theta_ep} for the angular dependence
is similar to the behavior of the lines for the $Q^2$ dependence in  Fig.~\ref{exp12}.

\begin{figure}[h!]
\centering
\hspace{-9mm}
\includegraphics[width=0.50\textwidth]{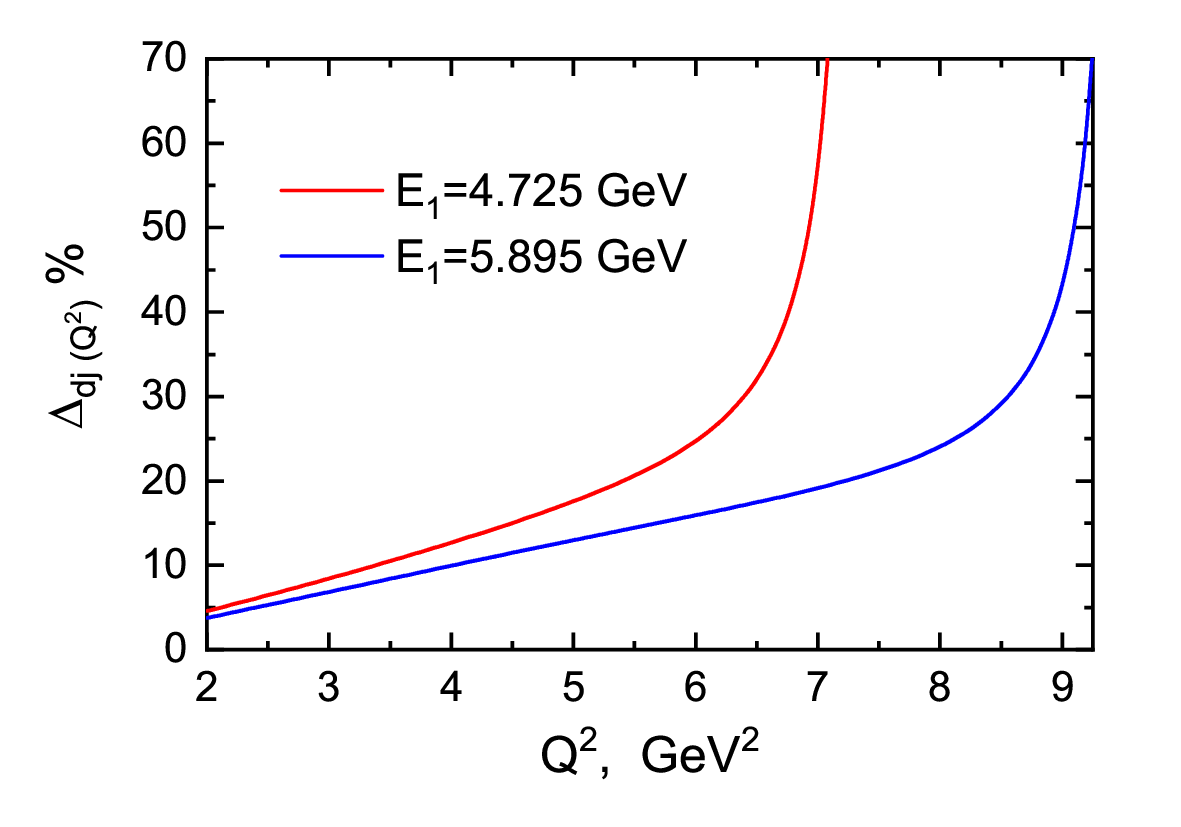}
\vspace{-5mm}
\caption{  
$Q^2$ dependence of the relative difference $\Delta_{dj}$ (\ref{Deltadj}) at electron
beam energies $E_1=4.725~\GEV$ (red line) and $E_1=5.895~\GEV$ (blue line).
For all lines, the degree of polarization of the proton target was taken to be the same $P_t=0.70$.
}
\label{exp14}
\end{figure}

Using the Kelly \cite{Kelly2004} and Qattan\cite{Qattan2015} parametrizations,
the relative difference $\Delta_{dj}$ between the polarization effects in the process
of $e \vec p \to \vec e p$  was calculated in the case of conservation and violation
of the scaling of the SFFs  as well as in the effects between these parameterizations $\Delta_{jk}$
\ba
\label{Deltadj}
\Delta_{dj}=\Big|\frac{\rm{Pd}-\rm{Pj}}{\rm{Pd}}\Big|, \;
\Delta_{jk}=\Big|\frac{\rm{Pj} - \rm{Pk}}{\rm{Pj}}\Big|,
\ea
where $P_d$, $P_j$, and $P_k$ are the polarizations calculated by formula~(\ref{lambdae}) for
$\lambda_{e_{2}}^{f}$ when using the corresponding parametrizations $R_d$, $R_j$, and $R_k$.
The results of calculations of $\Delta_{dj}$ at electron beam energies of
$4.725$ GeV and $5.895$ GeV are shown in Fig.~\ref{exp14}.

It follows from the graphs in Fig.~\ref{exp14} that the relative difference between the polarization
transferred from the initial proton to the final electron in the $e\vec p\to \vec e p$ process
in the case of conservation and violation of the scaling of the SFFs can reach 70\%, which can be used
to set up a polarization experiment by measuring the ratio $R$.

\begin{table*}[h!tpb]
\centering
\caption{
The degree of longitudinal polarization of the scattered electron $\lambda_{e_{2}}^{f}$ (\ref{lambdae})
at $E_1$ and $Q^2$ used in the experiment \cite{Liyanage2020}. The values in the columns
for $P_d$, $P_j$, and $P_k$ correspond to dipole dependence and
the  Qattan \cite{Qattan2015} and Kelly \cite{Kelly2004} parametrization (\ref{Rdj}). The corresponding electron
and proton scattering angles (in degrees) are given in columns for $\theta_{e}$ and $\theta_{p}$.
}
\vspace{1mm}
\label{DeltaKelly}
\tabcolsep=2.90mm
\footnotesize
\begin{tabular}
{| c | c | c | c | c | c | c | c | c |}
\hline
$E_1$, \rm{GeV}
&  $Q^2$, \rm{GeV}$^2$
& $\theta_{e}\, (^{\circ})$
& $\theta_{p} \,(^{\circ})$  
& $P_d$
& $P_j$
& $P_k$
& $\Delta_{dj}$, \%
& $\Delta_{jk}$, \% \\
\hline
5.895 &2.06 & 15.51 & 45.23 & --0.170 & --0.163 & --0.163 & 4.1 & 0.0 \\
\hline
5.895 &5.66 & 33.57 & 24.48 & --0.363 & --0.309 & --0.308 & 14.9  & 0.3 \\
\hline
4.725 &2.06 & 19.97 & 43.27 & --0.207 & --0.197 & --0.197 & 4.8 & 0.0 \\
\hline
4.725 &5.66 & 49.50 & 19.77 & --0.336 & --0.263 & --0.262 & 21.7 & 0.6\\
\hline
\end{tabular}
\end{table*}

Numerical values of the polarization transferred to the final electron in the $e\vec p\to\vec e p$
process for the three considered parametrizations of the ratio $R$ at $E_1$ and $Q^2$ used
in the experiment \cite{Liyanage2020}, are represented in Table~\ref{DeltaKelly}. In it, the columns
of values $P_d$, $P_j$, and $P_k$ correspond to the dipole dependence $R_d$, parametrizations
$R_j$ (\ref{Rdj}) from  \cite{Qattan2015} and $R_k$ \cite{Kelly2004};  the columns $\Delta_{dj}$, $\Delta_{jk}$ correspond
to the relative difference (\ref{Deltadj}) (expressed in percent) at electron beam energies
of $4.725$ GeV and $5.895$ GeV and two values of $Q^2$ equal to $2.06~\GEV^2$ and $5.66~\GEV^2$.
It follows from Table~\ref{DeltaKelly} that the relative difference between $Pj5$ and $Pd5$ at $Q^2=2.06~\GEV^2$
is 4.1\% and between $Pj4$ and $Pd4$ it is 4.8\%. At $Q^2 = 5.66~\GEV^2$, the difference increases
and becomes equal to 14.9 \% and 21.7\%, respectively. Note that the relative difference $\Delta_{jk}$
between $P_j$ and $P_k$ for all $E_1$ and $Q^2$ in Table~\ref{DeltaKelly} is less than 1\%.

\section{ Polarization effects in the  $\vec e \vec p \to  e p $  process}

\subsection{ Cross section of the $\vec e \vec p \to  e p $ process }

In the OPE approximation, the differential cross section of the 
process (\ref{EPEPe1p1}), calculated in an arbitrary reference
frame in DSB (\ref{DSBe1p1}), read
\ba
\label{T2alld}
\frac {d \sigma_{\vec e \vec p \to e p}}{d t}&=&\frac{ \pi \alpha^2 }{\lambda_s (1+\tau_p)}\,
\frac {|{T}|^2}{t^2} ,\\
\label{T4yd}
|T|^2& =& I_0 + \lambda_{e_{1}} \lambda_{p_{1}}  I_1, \\
I_0&=&G_E^{\,2} Y_1 + \tau_p\, G_M^{\,2} Y_2, \nn \\
I_1&=& \tau_p (G_E G_M Y_3 + G_M^{\,2} Y_4), \nn
\ea
where $\lambda_{e_{1}}$ and $\lambda_{p_{1}}$ are the degrees of polarization of the initial
electron and proton, and the functions  $Y_i$ ($i=1, \ldots 4$) read
\ba
Y_1&=&(p_+q_+)^2+q_+^2q_-^2, \nn \\
\label{Y2epd}
Y_{2}&=&(p_+ q_+)^2-q_+^2(q_-^2+4 m_0^2),\\
-Y_{3}&=&2\kappa_2\, m^2((p_+q_+)^2+q_+^2(q_-^2-4\,m_0^2\,))\,z_2^2,\nn \\
Y_{4}&=& 2 (m^2 p_+ q_+ -\kappa_2 q_+^2)(\kappa_2 p_+ q_+ - m_0^2 q_+^2) z_2^2,\nn \\
&&z_2=   
( \kappa_2^{2} - m^2 m_0^2 )^{-1/2}\,, \;\kappa_2=q_{1}p_{1} .\nn
\ea
In the case of arbitrary spin 4-vectors $s_{e_{1}}$ and $s_{p_{1}}$, the expressions
for $Y_3$ and $Y_4$ in the cross section (\ref{T2alld}) have the form:
\ba
\label{Y3aepd}
Y_{3}&=& 8\,m_0 m \, (s_{p_{1}})^{\mu}   
(g^{\bot})_{\mu \nu} (s_{e_{1}})^{\nu} q_+^2 \,,\\
Y_{4}&=& 8\,m_0 m \,(q_+ s_{p_{1}}) (q_+ s_{e_{1}}). \nn
\ea
Note, first, that the polarized part of the cross section (\ref{T2alld}) includes the term
with $Y_3$ 
containing the product of the SFFs, $G_E G_M$, and 
according to (\ref{Y3aepd}), is determined by the transverse part of the metric tensor
$(g^{\bot})_{\mu\nu}$ (\ref{gperp}).
Second, there is no similar structure in the cross section of the
$e\vec p\to e\vec p$ process in the DSB (\ref{DSBp1p2}), 
see \cite{JETPL2021,PEPAN2022,JETPL2022}.

In the ultrarelativistic limit, when the mass of an electron can be neglected,
for the functions $Y_i$ (\ref{Y2epd}) in the LF 
we obtain expressions, that depend only on the energy of the initial and final electrons:
\ba
Y_1 &=&8m^2 (2 E_1E_2 - m E_- ), \nn \\
Y_2&=&8 m^2 (E_1^{\,2} + E_2^{\,2} + m E_-),\\
\label{Y2d}
-Y_3&=&(2m/E_1)\, Y_1 ,\nn \\
-Y_4&=&8 m^2 E_+ E_- (m+E_1)/E_1.\nn
\ea
The polarization asymmetry in the process (\ref{EPEPe1p1}) is determined by the square
of the amplitude modulus (\ref{T4yd}) as follows \cite{Alguard1976,Alguard1976b}:
\ba
A=\frac {|{T}|^2 (\lambda_{e_{1}}=-1) - |{T}|^2 (\lambda_{e_{1}}=+1) }
{ |{T}|^2 (\lambda_{e_{1}}=-1)  + |{T}|^2(\lambda_{e_{1}}=+1) }. \nn
\ea
As a result we have
\ba
&& A=-\lambda_{p_{1}}
\frac{\tau_p \, (G_E G_M Y_3 + G_M^{\,2} Y_4)}{G_E^{\,2} Y_1 + \tau_p G_M^{\,2} Y_2}. \nn
\ea
By dividing the numerator and denominator in the last expression into $Y_1 G_M^{\,2}$ and introducing
the experimentally measured ratio $R \equiv \mu_p G_E/G_M$, we get:
\ba
\label{lambdaea}
A=-\lambda_{p_{1}}\, \frac{\mu_p \tau_p\, ( (Y_3/Y_1) R+ \mu_p (Y_4/Y_1))}
{ R^{\,2} + \mu_p^2 \,\tau_p\, (Y_2/Y_1) \,}.
\ea
Note that  expressions (\ref{lambdaea}) for polarization asymmetry in the $\vec e \vec p \to e p$
process and the expressions (\ref{lambdae}) for electron transferred polarization $\lambda_{e_{2}}^{f}$
in the $e \vec p \to \vec e p$ process coincide up to the sign.
For this reason, the quadratic equation for extracting the ratio
$R$  coincides with the explicit form of   equation (\ref{Rp}) and has
coefficients of the same shape, except for one: $\alpha_0=-A_{exp} / \lambda_{p_{1}}$,
where $A_{exp}$ is experimentally measured polarization asymmetry.
%
\begin{figure}[h!]
\centering
\includegraphics[width=0.50\textwidth]{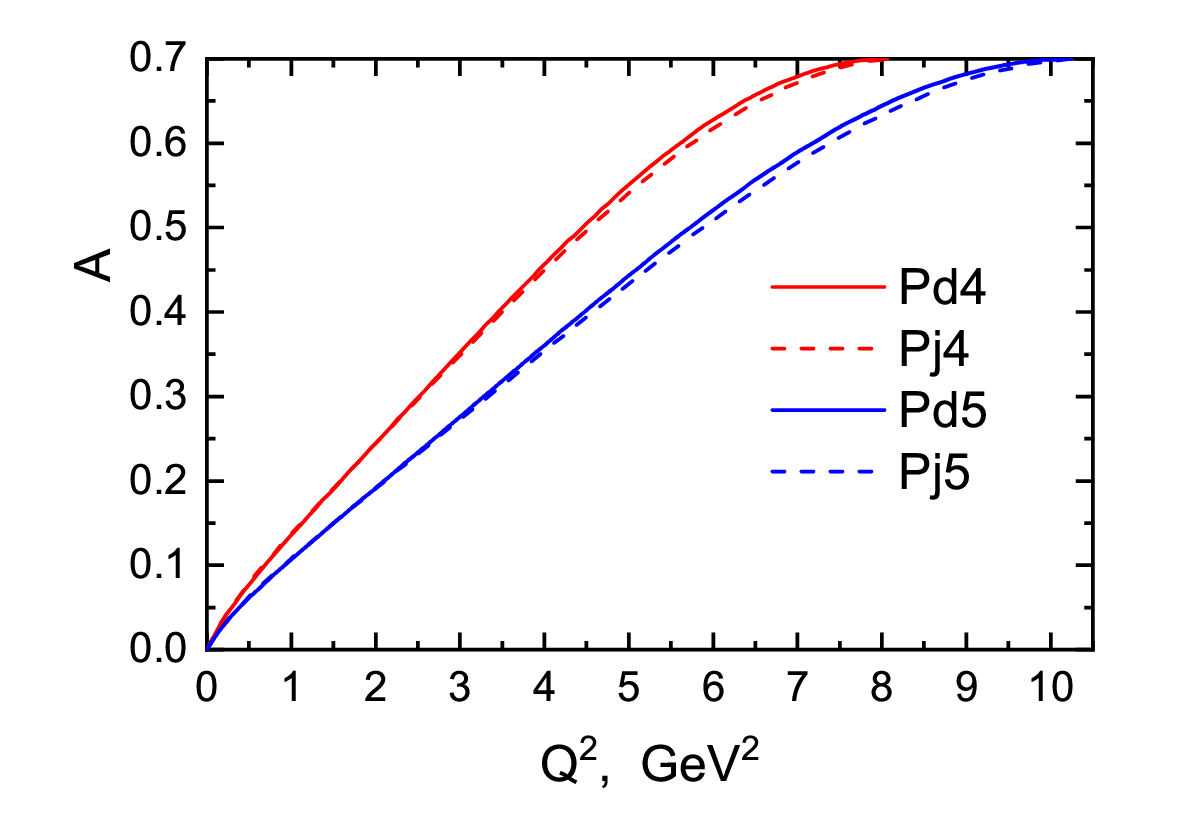}
\vspace{-8mm}
\caption{ 
$Q^2$ dependence of the polarization asymmetry of $A$ (\ref{lambdaea}) in the
$\vec e \vec p \to e p$ process at electron beam energies in the experiment \cite{Liyanage2020}.
The lines $Pd4, Pd5$ (solid) and $Pj4, Pj5$ (dashed) correspond to the dipole dependence
$R=R_d$ ($R_d=1$) and parametrization of $R=R_j$ (\ref{Rdj}) from \cite{Qattan2015}.
The lines $Pd4, Pj4$ ($Pd5, Pj5$) correspond to the electron beam energy $4.725$ GeV and $5.895$ GeV.
For all lines, the degree of polarization of the proton target was taken to be the same $P_t=0.70$.
}
\label{asimme1p1}
\end{figure}

The results of  numerical calculations of the $Q^2$ dependence of the polarization
asymmetry $A$ (\ref{lambdaea}) in the $\vec e \vec p \to e p$ process 
at electron beam energies $E_1=4.725$ GeV and 5.895 GeV are represented by graphs
in the Fig. \ref{asimme1p1}, from which it follows that this dependence
for each of the energies of the electron beam is almost linear.
With   increasing of $Q^2$ from 0 to $Q^2_{max}$, it changes from 0 to $P_t=0.70$ at the boundaries
of the spectrum at $Q^2=Q^2_{max}$. The effects of   scaling violations are small in
the entire range of acceptable values of $Q^2$; they do not exceed 1.79 \% at $E_1=4.725$ GeV
and 2.32 \% at $E_1=5.895$ GeV. For this reason, the measurement of polarization asymmetry in
the $ \vec e \vec p \to e p $ process can be used as a test
to verify the conservation of the SFF scaling.

Note that the double spin asymmetry in the elastic process $\vec e \vec p \to ep$
in the case when the spin quantization axes of a resting proton target and an incident electron beam
are parallel was first measured in the experiment \cite{Alguard1976}, as a result of which it was
first established that the SFFs ratio $R$ is positive.

\section{Conclusion}

In this paper, in the one-photon exchange approximation we analyze polarization effects
in the elastic $\vec e \vec p \to e p$ and $ e \vec p \to \vec e p$ processes in the case
when the spin quantization axes of the target proton at rest and the incident or scattered
electron are parallel.
To do this, in the kinematics of the SANE Collaboration experiment \cite{Liyanage2020}
using the Kelly \cite{Kelly2004} and Qattan \cite{Qattan2015} parametrizations for the Sachs form factor
ratio $R\equiv \mu_p G_E/G_M$, a numerical analysis was carried out of the dependence of
the longitudinal polarization degree transferred to the scattered electron in the $e \vec p \to \vec e p$
process and the double spin asymmetry in the $\vec e \vec p \to e p$ process on the square
of the momentum transferred to the proton as well as on the scattering angle of the electron.
As it turned out, the Kelly \cite{Kelly2004} and Qattan \cite{Qattan2015} parametrizations
give almost identical results. 

As a result of calculations, it was established that the relative difference
in the longitudinal polarization degree of the final electron in the
$ e \vec p \to \vec e  p$ process for the case of conservation and violation
of the SFF scaling can reach 70 \%, which can be used to conduct a polarization experiment of
a new type of measurement of the SFF ratio $R$.

For the double spin asymmetry in the $\vec e \vec p \to e p$  process this difference
is rather small and does not exceed 1.79 \% for the electron beam energy $E_1=4.725$ GeV
and 2.32 \% for $E_1=5.895$ GeV. For this reason, the measurement of polarization asymmetry
in the $ \vec e \vec p \to e p $ process can be used as a test to verify the conservation of the SFF scaling.

At present, the experiment on measuring the degree of longitudinal polarization transferred
to the final electron in the process $e \vec p \to \vec e p$ seems to be quite realistic,
since a proton target with a high degree of polarization $P_t = 70 \pm 5$ \% has been already created
and is used in the experiment \cite{Liyanage2020}. For this
reason, it would be most appropriate to conduct the proposed experiment at the setup
used in \cite{Liyanage2020} at the same proton polarization degree $P_t=0.70$, electron beam
energies $E_1=4.725$ GeV and 5.895 GeV.

The difference between the proposed experiment and the one in \cite{Liyanage2020}
consists in the fact that an incident electron beam must be unpolarized, and the detected
scattered electron must move strictly along the direction of the spin quantization axis
of a resting proton target. In the proposed experiment, it is necessary to measure only
the longitudinal polarization degree of the scattered electron,which is an advantage
compared to the AR method \cite{Rekalo74} used in JLab polarization experiments.

\section*{ACKNOWLEDGMENTS}
This work was carried out within the framework of Belarus-JINR  scientific cooperation
 and State Program of Scientific Research ``Convergence-2025'' of the Republic of Belarus
under Projects No.~20241529 and No.~20210852.


\begin{thebibliography}{60}

\bibitem{Hofstadter1958}
R. Hofstadter, F. Bumiller, and M. R. Yearian, 
\href{https://link.aps.org/doi/10.1103/RevModPhys.30.482}{Rev. Mod. Phys. {\bf 30}, 482 (1958)}.

\bibitem{Rosen} M.\,N.\, Rosenbluth,
\href{https://doi.org/10.1103/PhysRev.79.615}{Phys. Rev. {\bf 79}, 615 (1950)}.

\bibitem{Dombey} N.~ Dombey,
\href{https://doi.org/10.1103/RevModPhys.41.236}{Rev. Mod. Phys. {\bf 41}, 236 (1969)}.

\bibitem{Rekalo74}
A. I.~Akhiezer and M. P.~Rekalo, Sov. J. Part. Nucl. {\bf 4}, 277 (1974);
Fiz. Elem. Chast. Atom. Yadra {\bf 4}, 662 (1973). 

\bibitem{AR}
A.\,I.\, Akhiezer and M.\,P.\, Rekalo, {\it Electrodynamics of Hadrons}
(Naukova Dumka, Kiev, 1977) [in Russian].

\bibitem{GL97}
M.\, V.\, Galynskii and M.\, I.\, Levchuk,
Phys. Atom. Nucl. {\bf 60}, 1855 (1997); Yad. Fiz. {\bf 60}, 2028 (1997).

\bibitem{ETG15}  S.~Pacetti, R.~Baldini Ferroli, and E.~Tomasi-Gustafsson,
\href{ https://doi.org/10.1016/j.physrep.2014.09.005}{Phys. Rept. {\bf 550-551}, 1 (2015)}.

\bibitem{Miller2015}
Y. S. Liu and G. A. Miller,
\href{https://doi.org/10.1103/PhysRevC.92.035209}{Phys. Rev. C {\bf 92},  035209 (2015)}.

\bibitem{Jones00} M.~K.~Jones, K.~A.~Aniol, F.~T.~Baker \textit{et al.},
\href{https://dx.doi.org/10.1103/PhysRevLett.84.1398}{Phys. Rev. Lett. {\bf 84},  1398 (2000)}.

\bibitem{Gay01}
O. Gayou, K. Wijesooriya, A. Afanasev {\it et al.},
\href{https://dx.doi.org/10.1103/PhysRevC.64.038202}{Phys.\ Rev.\ C {\bf 64}, 038202 (2001)}.

\bibitem{Gay02}
O. Gayou, K. A. Aniol, T. Averett {\it et al.},
\href{https://dx.doi.org/10.1103/PhysRevLett.88.092301}{Phys.\ Rev.\ Lett. {\bf 88}, 092301 (2002)}.

\bibitem{Pun05} V.~ Punjabi, C.F. Perdrisat,  K.A. Aniol {\it et al.},
\href{https://doi.org/10.1103/PhysRevC.71.055202}{Phys. Rev. C {\bf 71}, 055202 (2005)}.

\bibitem {Puckett10} A. J. R. Puckett, E. J. Brash, M. K.Jones {\it et al.},
\href{https://doi.org/10.1103/PhysRevLett.104.242301}{Phys.\ Rev.\ Lett.\ {\bf 104}, 242301 (2010)}.

\bibitem {Puckett12}
A. J. R. Puckett, E. J. Brash, O. Gayou {\it et al.},
\href{https://doi.org/10.1103/PhysRevC.85.045203}{Phys. Rev. C {\bf 85}, 045203 (2012).}

\bibitem {Puckett17} A. J. R. Puckett,  E. J. Brash, M. K. Jones {\it et al.},
\href{https://doi.org/10.1103/PhysRevC.96.055203}{Phys. Rev. C {\bf 96}, 055203 (2017).}

\bibitem{Qattan2005}
I. A. Qattan, J. Arrington, R. E. Segel {\it et al.},
\href{https://doi.org/10.1103/PhysRevLett.94.142301}{Phys. Rev. Lett. {\bf 94},  142301 (2005).}

\bibitem {Liyanage2020}
A. Liyanage {\it et al.} (SANE Collaboration), 
\href{https://doi.org/10.1103/PhysRevC.101.035206}{Phys.\ Rev.\ C {\bf 101}, 035206 (2020)}.

\bibitem {Donnelly1986}
T. W. Donnelly and A. S. Raskin,
\href{https://doi.org/10.1016/0003-4916(86)90173-9}{Ann. Phys. {\bf 169}, 247 (1986)}.
\bibitem {Bernauer1}
J. C. Bernauer {\it et al.},
\href{https://dx.doi.org/10.1103/PhysRevLett.105.242001}{Phys. Rev. Lett. {\bf105}, 242001 (2010)}.

\bibitem {Bernauer2}
J. C. Bernauer {\it et al.},
\href{https://dx.doi.org/10.1103/PhysRevC.90.015206}{Phys. Rev. C {\bf90}, 015206 (2014)}.

\bibitem {Bernauer3}
J.C. Bernauer,
\href{ http://inspirehep.net/record/1358265/}{PhD Thesis, Mainz Universitat, Institute for Kernphysics}.

\bibitem{Arrington:2011dn}
J.~Arrington, P.~G.~Blunden, and W.~Melnitchouk,
\href{https://doi:10.1016/j.ppnp.2011.07.003}{Prog. Part. Nucl. Phys. \textbf{66}, 782 (2011)}.

\bibitem{Punjabi2015}
V.\, Punjabi, C.\, F.\, Perdrisat, M.\, K.\, Jones, E.\, J.\, Brash, and C.\, E.\, Carlson,
\href{https://doi.org/10.1140/epja/i2015-15079-x}{Eur.\,Phys.\,J.\,A {\bf 51}, 79 (2015)}.

\bibitem{Afanasev2017}
A. Afanasev, P.G. Blunden, D. Hasell, B.A. Raue,
\href{https://doi.org/10.1016/j.ppnp.2017.03.004}{Prog. Part. Nucl. Phys. {\bf 95}, 245278 (2017)}.

\bibitem{Rachek:2014fam}
I.~A.~Rachek \textit{et al.} (VEPP-3 Collaboration), 
\href{https://doi:10.1103/PhysRevLett.114.062005}{Phys. Rev. Lett. \textbf{114}, 062005 (2015)}.

\bibitem{OLYMPUS:2016gso}
B.~S.~Henderson \textit{et al.} (OLYMPUS Collaboration),
\href{https://doi:10.1103/PhysRevLett.118.092501}{Phys. Rev. Lett. \textbf{118}, 092501 (2017)}.

\bibitem{CLAS:2013mza}
M.~Moteabbed \textit{et al.} (CLAS Collaboration),
\href{https://doi:10.1103/PhysRevC.88.025210}{Phys. Rev. C \textbf{88}, 025210  (2013)}.

\bibitem{CLAS:2014xso}
D.~Adikaram \textit{et al.} (CLAS Collaboration),
\href{https://doi:10.1103/PhysRevLett.114.062003}{Phys. Rev. Lett. \textbf{114}, 062003 (2015)}.

\bibitem{CLAS:2016fvy}
D.~Rimal \textit{et al.} (CLAS Collaboration),
\href{https://doi:10.1103/PhysRevC.95.065201}{Phys. Rev. C \textbf{95}, 065201  (2017)}.

\bibitem{Bernauer2024}
R. Alarcon, R. Beck, J. C. Bernauer \textit{et al.},
\href{https://doi.org/10.1140/epja/s10050-024-01299-2}{Eur. Phys. J. A  {\bf 60}, 81 (2024)}.

\bibitem{JETPL2008}
M. V. Galynskii, E. A. Kuraev, and Yu. M. Bystritskiy,
\href{https://doi.org/10.1134/S0021364008200034}{JETP Lett. {\bf 88}, 481 (2008)}.

\bibitem{JETPL18}  M. V. Galynskii,
\href{https://doi.org/10.1134/S0021364019010089}{JETP Lett. {\bf 109}, 1 (2019)}.

\bibitem{JETPL19}
M. V.  Galynskii and R. E. Gerasimov,
\href{https://doi.org/10.1134/S0021364019220077}{JETP Lett. {\bf 110}, 646 (2019)}.

\bibitem{JETPL2021} M. V. Galynskii,
\href{https://doi.org/10.1134/S0021364021090083}{JETP Lett. {\bf 113}, 555 (2021)}.

\bibitem{PEPAN2022} M. V. Galynskii,
\href{https://doi.org/10.1134/S1547477122010058}{Phys. Part. Nucl. Lett. {\bf 19}, 26 (2022)}.

\bibitem{JETPL2022}
M. V. Galynskii, 
\href{https://doi.org/10.1134/S0021364022601804}{JETP Lett. {\bf 116}, 420 (2022)}.

\bibitem{Kelly2004} J. J. Kelly,
\href{https://dx.doi.org/10.1103/PhysRevC.70.068202 }{Phys. Rev. C {\bf 70}, 068202 (2004)}.

\bibitem{Qattan2015}
I. A. Qattan, J. Arrington, and A. Alsaad, 
\href{https://dx.doi.org/10.1103/PhysRevC.91.065203 }{Phys. Rev. C {\bf 91}, 065203 (2015)}.

\bibitem{PRD2023}
M. V. Galynskii, Yu. M. Bystritskiy, and V. M. Galynsky,
\href{https://doi.org/10.1103/PhysRevD.108.096032}{Phys. Rev. D {\bf 108}, 096032 (2023)}.

\bibitem {Brash1}
E. J. Brash, A. Kozlov, S. Li, G. M. Huber,
\href{https://dx.doi.org/10.1103/PhysRevC.65.051001}{Phys. Rev. C {\bf65}, 051001 (2002)}.

\bibitem {Graczyk2}
K. M. Graczyk, P. Plonski, R. Sulej,
\href{https://dx.doi.org/10.1007/JHEP09(2010)053}{JHEP {\bf09}, 053 (2010)}.

\bibitem {Sufian3}
R. S. Sufian, G. F. de Teramond, S. J. Brodsky, A. Deur, H. G. Dosch,
\href{https://dx.doi.org/10.1103/PhysRevD.95.014011}{Phys. Rev. D {\bf95}, 014011 (2017)}.

\bibitem {Alberico6}
W. M. Alberico, S. M. Bilenky, C. Giunti, K. M. Graczyk,
\href{https://dx.doi.org/10.1103/PhysRevC.79.065204}{Phys. Rev. C {\bf79},  065204 (2009)}.

\bibitem {Borah7}
Kaushik Borah, Richard J. Hill, Gabriel Lee, and Oleksandr Tomalak
\href{https://dx.doi.org/10.1103/PhysRevD.102.074012}{Phys. Rev. D {\bf102}, 074012 (2020)}.

\bibitem {Arrington8}
Z. Ye, J. Arrington, R. J. Hill, and G. Lee,
\href{https://dx.doi.org/10.1016/j.physletb.2017.11.023}{Phys. Lett. B {\bf777}, 8 (2018)}.

\bibitem{Jacob} M.\, Jacob and G.\, Wick,
\href{https://dx.doi.org/10.1016/0003-4916(59)90051-X}{Ann. Phys. {\bf 7}, 404 (1959)}.

\bibitem{FIF70} F.\,I.\, Fedorov,
\href{https://dx.doi.org/10.1007/BF01038044}{Theor. Math. Phys. {\bf 2}, 248 (1970)}.

\bibitem{GL}
F.\,I.\, Fedorov, {\it The Lorentz Group} (Nauka, Moscow, 1979) [in Russian].

\bibitem{AB}
A. I. Akhiezer and V. B. Berestetskii, {\it Quantum Electrodynamics}, 3rd ed.
(Nauka, Moscow, 1969; Wiley, New York, 1965).

\bibitem{BLP}
V. B. Berestetskii, E. M. Lifshits, and L. P. Pitaevskii, {\it Course of Theoretical Physics}, Vol. 4:
{\it Quantum Electrodynamics} (Nauka, Moscow, 1989; Pergamon, Oxford, 1982).

\bibitem{Sik84}
S. M. Sikach, Vesti Akad. Nauk BSSR, Ser. Fiz. Mat. Nauk, {\bf 2}, 84 (1984) [in Russian].

\bibitem{GS98}
M.\,V.\, Galynskii, S.\,M.\, Sikach, 
\href{https://sci-hub.ru/10.1134/1.953087}{Phys. Part. Nucl. {\bf 29}, 469 (1998);
Fiz. Elem. Chast. Atom. Yadra {\bf 29}, 1133 (1998)}.

\bibitem{Tomalak:2014dja}
O.~Tomalak and M.~Vanderhaeghen,
\href{https://doi:10.1103/PhysRevD.90.013006}{Phys. Rev. D \textbf{90}, 013006 (2014)}.

\bibitem{Gakh2008}
G.~I.~Gakh, E.~Tomasi-Gustafsson,
\href{https://doi.org/10.1016/j.nuclphysa.2007.11.001}{Nucl. Phys. A {\bf 799}, 127 (2008)}.

\bibitem{Alguard1976}
M.~J.~ Alguard {\it et al.}
\href{https://doi.org/10.1103/PhysRevLett.37.1258}{Phys.~Rev.~Lett. {\bf 37}, 1258 (1976)}.

\bibitem{Alguard1976b}
M.~J.~ Alguard {\it et al.}
\href{https://doi.org/10.1103/PhysRevLett.37.1261}{Phys.~Rev.~Lett. {\bf 37}, 1261 (1976)}.



\bibitem{Tomalak:2018jak}
O.~Tomalak and M.~Vanderhaeghen,
\href{https://doi:10.1140/epjc/s10052-018-5988-5}{Eur. Phys. J. C \textbf{78}, 514 (2018)}.

\bibitem{Korchagin2021}
 N.~Korchagin, and A.~Radzhabov,
\href{https://arxiv.org/abs/2106.06883}{arXiv: 2106.06883 [nucl-th]}.

\bibitem{Weiss2023}
 F.~Gil-Dominguez, J.~Alarcon, C.~Weiss,
\href{https://doi.org/10.1103/PhysRevD.108.074026}{Phys. Rev. D {\bf 108}, 074026 (2023)}.

\bibitem{Bernauer2021}
Ethan Cline, Jan C. Bernauer, Axel Schmidt,
\href{https://doi.org/10.1140/epja/s10050-021-00597-3}{Eur. Phys. J. A \textbf{57}, 290 (2021)}.


\end{thebibliography}
\end{document}